Final draft manuscript
with a post-publication correction to Eq. 35
(the erratum for this correction has been incorporated into this version of the manuscript)

# Thermal Extraction of Volatiles from Lunar and Asteroid Regolith in Axisymmetric Crank-Nicholson Modeling

Philip T. Metzger, Ph.D. (1), Kris Zacny, Ph.D. (2), and Phillip Morrison (3)

(1) Florida Space Institute, University of Central Florida, 12354 Research Parkway, Partnership 1 Building, Suite 214, Orlando, FL 32826-0650; PH (407) 823-5540; email: philip.metzger@ucf.edu
(2) Honeybee Robotics, 398 W Washington Blvd suite 200, Pasadena, CA 91103; email: KAZacny@honeybeerobotics.com
(3) Honeybee Robotics, 398 W Washington Blvd suite 200, Pasadena, CA 91103; email: PDMorrison@honeybeerobotics.com

## ABSTRACT

A physics-based computer model has been developed to support the development of volatile extraction from regolith of the Moon and asteroids. The model is based upon empirical data sets for extraterrestrial soils and simulants, including thermal conductivity of regolith and mixed composition ice, heat capacity of soil and mixed composition ice, hydrated mineral volatile release patterns, and sublimation of ice. A new thermal conductivity relationship is derived that generalizes cases of regolith with varying temperature, soil porosity, and pore vapor pressure. Ice composition is based upon measurements of icy ejecta from the Lunar CRater Observation and Sensing Satellite (LCROSS) impact and it is shown that thermal conductivity and heat capacity equations for water ice provide adequate accuracy at the present level of development. The heat diffusion equations are integrated with gas diffusion equations using multiple adaptive timesteps. The entire model is placed into a Crank-Nicholson framework where the finite difference formalism was extended to two dimensions in axisymmetry. The one-dimensional version of the model successfully predicts heat transfer that matches lunar and asteroid data sets. The axisymmetric model has been used to study heat dissipation around lunar drills and water extraction in asteroid coring devices.

## INTRODUCTION

There is growing interest in extracting water from the Moon (Casanova, et al., 2017), from Mars (Abbud-Madrid et al., 2016), and from asteroids (Nomura et al., 2017). NASA's Lunar CRater Observation and Sensing Satellite (LCROSS) demonstrated the existence of water ice on the Moon when it impacted into Cabeus crater, a permanently shadowed region (PSR) near the Moon's south pole. The resulting ejecta blanket was found to contain water and other volatiles (Colaprete et al., 2010; Gladstone et al., 2010). Carbonaceous asteroids contain hydrated and hydroxylated phyllosilicates (Jewitt et al., 2007). These volatiles are stable at typical asteroid temperatures in near Earth orbits in vacuum, but the water evolves when heated to moderate temperature (Zacny et al., 2016). Mars has abundant water in the form of glacial deposits, hydrated minerals including polyhydrated sulfate minerals and hydrated phyllosilicates, and water adsorbed globally at a low weight percent onto the surfaces of regolith grains (Abbud-Madrid et al., 2016).

Extraction of water on the Moon or Mars can be accomplished through strip mining with subsequent processing of the mined material or through in situ thermal techniques: injecting heat into the subsurface, providing a means for vapor or liquid to reach the surface, and collecting it in tanks. Methods on asteroids could be similar or could involve bagging the entire asteroid and heating or spallation of the rocky material with concentrated sunlight (Sercel et al., 2016). Water can be used to make rocket propellant to reduce the cost of operating in space (Sanders, et al., 2008; Hubbard, et al., 2013; Sowers, 2016; Kutter and Sowers, 2016), can serve as passive radiation shielding for astronauts (Parker, 20016), and can provide life support (Kelsey et al., 2013). In addition to supporting national space agency activities, water-derived rocket propellant can be used commercially for boosting telecommunication satellites from low Earth orbit or from geosynchronous transfer orbit into geostationary orbit, or for supporting space tourism or other non-governmental activities (Metzger, 2016).

It is difficult to test extraterrestrial water extraction technologies on Earth because of the high preparation costs for realistic test environments: large-scale beds of frozen regolith in vacuum or Mars atmosphere chambers (Kleinhenz and Linne, 2013). Nevertheless, testing is vital, as shown by Zacny et al. (2016) performing thermal extraction of water from icy lunar simulant. Those tests found that water vapor moves through the regolith down the thermal gradient away from the hot mining device, so depending on the particular geometry of the system it could either collect large amounts of water or none at all. This is highly dependent on environmental conditions including vacuum, ice characteristics, and thermal state, so testing without the correct environment would have little value. However, the environments can be simulated numerically to perform low-cost digital design evaluation in lieu of some of the testing, reducing the cost and speeding the schedule of hardware development. For this approach to work, the equations must accurately predict the thermodynamic behaviors of icy, extraterrestrial soil or hydrated minerals in hard vacuum or in low pressure conditions and at extreme temperatures. This is still challenging because most measurements of thermal conductivity for soils has been performed in terrestrial conditions with liquid water content, Earth's atmospheric pressure, and/or ambient temperatures. Therefore, more work is needed developing improved models.

The application that led to the present effort is the World Is Not Enough (WINE) spacecraft concept, which is being developed by Honeybee Robotics under NASA contract (Zacny et al., 2016). WINE will be a small spacecraft, approximately 27U in CubeSat dimensions (3 by 3 by 3 cubes), with legs for walking short distances and a steam propulsion system for hopping multiple kilometers (Metzger et al., 2016). WINE spacecraft could operate on a body such as dwarf planet Ceres obtaining water from hydrated minerals that may exist on its surface, or on a moon like Europa where ice is abundant. WINE will drive a corer into the regolith to extract water and perform science and prospecting measurements on the regolith. The water will be extracted thermally by heating the material in the corer. Vapor will travel into a collection chamber where it is frozen onto a cold finger. After multiple coring operations have collected enough water, the tank will be heated to high pressure and vented through a nozzle to produce hopping thrust. Development and validation of the coring and water extraction system requires at least 2D (axisymmetric) computer modeling of heat transfer in regolith. The modeling is needed to determine energy requirements for this process to set requirements for the spacecraft power system and to determine whether solar energy is adequate or whether Radioisotopic Heater Units (RHUs) are needed to generate adequate thermal energy on a particular planet.

The authors were team members for another application of this modeling: NASA's Resource Prospector mission (which was cancelled while in development). It was planned to prospect for water in the Moon's polar regions by drilling into the regolith, bringing up cuttings for physical and chemical analysis. One objective of the mission was to determine the temperature of the subsurface regolith around the drill sites. Unfortunately, drilling creates a lot of heat and experiments showed that it takes hours or even days for the soil to cool back to the original temperature. The mission's timeline cannot afford for the rover to sit so long in one location waiting to take a measurement. Modeling may be able to help solve this problem, too. The cooling rate around the drill bit should depend on the boundary conditions, which in cylindrical coordinates centered on the drill is the ice temperature asymptotically far from the drill. If the natural subsurface temperatures are relatively constant over distances comparable to the radius that was heated by drilling, then the asymptotic temperature will equal the original temperature at the drilling location. Therefore, measuring only the cooling rate at one or several depths down the drill bit should be adequate for modeling to determine the original temperature of the subsurface. The model will need to be populated with information obtained from the drill cuttings, including density of the soil and ice content as a function of depth, including possibly chemistry of the ice as measured by the rover's instruments. The measured drill torque may contribute to calculating the original density of the regolith with depth. If the model has accurate constitutive equations, then with these measurements as inputs the model can be run repeatedly using different boundary conditions until it correctly reproduces the measured cooling rates around the drill bit. This concept needs to be developed through modeling, which requires improving the model's constitutive equations, followed by comparison to ground testing before the mission.

Mitchel and dePater (1994) developed a one-dimensional model of heat transfer for Mercury and the Moon. It included solar insolation at the surface, geophysical heat flux from the subsurface, and constitutive equations for heat capacity and thermal conductivity of the regolith. The model used a finite difference framework with the Crank-Nicolson algorithm. Vasavada et al. (1999) extended the model. Vasavada et al. (2012) compared the model to radiometer data of the Moon's surface heating and cooling throughout a lunar day. Hayne et al. (2017) used it to map apparent looseness of the lunar soil globally.

Here, the model methodology is extended in three ways. This extends the progress first reported by Metzger (2018). First, the model is converted into axisymmetric 2D Crank-Nicolson form. Second, the constitutive equations are extended based upon additional data sets for soil and mixed composition ice over varying temperatures, porosities, and gas pore pressures. Third, the heat transfer model is merged with algorithms for gas diffusion following the methodology of Scott and Ko (1968). Another model of heat and mass transfer for extraction volatiles from regolith was recently developed by Reiss (2018) using a different methodology than the one that is followed here, so comparing the two models in future work will provide a useful test of the methodologies.

## 2D Axisymmetric Crank Nicholson

The 1D thermal model described above has been reproduced and extended to 2D axisymmetric form. The 2D heat flux equation in Cartesian coordinates is,

$$\frac{\partial T}{\partial t} = \frac{k}{2\rho C}\left(\frac{\partial^2 T}{\partial x^2} + \frac{\partial^2 T}{\partial y^2}\right) \tag{1}$$

where $T$ is temperature, $k$ is thermal conductivity of the material, $\rho$ is density of the material, $C$ is heat capacity of the material, and $t$ is time. The equation is discretized for use in a finite difference model. The left-hand size of the discretized equation calculates the change in $T$ from before to after one time step. The right-hand side could therefore be evaluated either before or after that time step. The Crank-Nicolson method is simply to average these two approaches (Crank and Nicolson, 1947). This results in a linear system of equations that is stable and can be solved quickly. Using Crank-Nicholson discretization in cartesian coordinates with $\Delta y = \Delta x$, Eq. (1) becomes,

$$\frac{2(\Delta x)^2 \rho_i C_i^n}{\Delta t}\left(T_{ij}^{n+1} - T_{ij}^n\right) = k_{i-,j}^n\left(T_{i-1,j}^n + T_{i-1,j}^{n+1}\right) - \left(k_{i-,j}^n + k_{i+,j}^n\right)\left(T_{i,j}^n + T_{i,j}^{n+1}\right) + k_{i+,j}^n\left(T_{i+1,j}^n + T_{i+1,j}^{n+1}\right) + k_{i,j-}^n\left(T_{i,j-1}^n + T_{i,j-1}^{n+1}\right) - \left(k_{i,j-}^n + k_{i,j+}^n\right)\left(T_{i,j}^n + T_{i,j}^{n+1}\right) + k_{i,j+}^n\left(T_{i,j+1}^n + T_{i,j+1}^{n+1}\right) \quad (2)$$

Converting to axisymmetric form requires the extra terms in the radial derivative, so in cylindrical coordinates with $\Delta r = \Delta z$ it becomes,

$$\frac{2(\Delta z)^2 \rho_i C_i^n}{\Delta t}\left(T_{ij}^{n+1} - T_{ij}^n\right) = k_{i-,j}^n\left(T_{i-1,j}^n + T_{i-1,j}^{n+1}\right) - \left(k_{i-,j}^n + k_{i+,j}^n\right)\left(T_{i,j}^n + T_{i,j}^{n+1}\right) + k_{i+,j}^n\left(T_{i+1,j}^n + T_{i+1,j}^{n+1}\right) + k_{i,j-}^n\left(T_{i,j-1}^n + T_{i,j-1}^{n+1}\right) - \left(k_{i,j-}^n + k_{i,j+}^n\right)\left(T_{i,j}^n + T_{i,j}^{n+1}\right) + k_{i,j+}^n\left(T_{i,j+1}^n + T_{i,j+1}^{n+1}\right) + \frac{k_{i,j}^n\left[\left(T_{i,j+1}^n + T_{i,j+1}^{n+1}\right) - \left(T_{i,j-1}^n + T_{i,j-1}^{n+1}\right)\right]}{2j} \quad (3)$$

where the radial and vertical directions are $r$ and $z$, respectively, and the discretized radial distance is $r_j = j\Delta r = j\Delta z$. Collecting terms with $\alpha = k\Delta t/[2(\Delta z)^2 \rho C]$,

$$\left(1 + \alpha_{i-,j}^n + \alpha_{i+,j}^n + \alpha_{i,j-}^n + \alpha_{i,j+}^n\right)T_{ij}^{n+1} - \alpha_{i+,j}^n T_{i+1,j}^{n+1} - \alpha_{i-,j}^n T_{i-1,j}^{n+1} - \left(\alpha_{i,j-}^n - \frac{\alpha_{i,j}^n}{2j}\right)T_{i,j-1}^{n+1} - \left(\alpha_{i,j+}^n + \frac{\alpha_{i,j}^n}{2j}\right)T_{i,j+1}^{n+1} = \left(1 - \alpha_{i-,j}^n - \alpha_{i+,j}^n - \alpha_{i,j-}^n - \alpha_{i,j+}^n\right)T_{ij}^n + \alpha_{i+,j}^n T_{i+1,j}^n + \alpha_{i-,j}^n T_{i-1,j}^n + \left(\alpha_{i,j-}^n - \frac{\alpha_{i,j}^n}{2j}\right)T_{i,j-1}^n` + \left(\alpha_{i,j+}^n + \frac{\alpha_{i,j}^n}{2j}\right)T_{i,j+1}^n \quad (4)$$

Adapting the method of Summers (2012) to the axisymmetric case, two operators are defined as

$$\delta_z^2 T_{ij} = -\alpha_{i-,j}T_{i-1,j} + \left(\alpha_{i-,j} + \alpha_{i+,j}\right)T_{i,j} - \alpha_{i+,j}T_{i+1,j} \quad (5)$$

and

$$\delta_r^2 T_{ij} = \left(\alpha_{i,j-} - \frac{\alpha_{i,j}}{2j}\right)T_{i,j-1} - \left(\alpha_{i,j-} + \alpha_{i,j+}\right)T_{i,j} + \left(\alpha_{i,j+} + \frac{\alpha_{i,j}}{2j}\right)T_{i,j+1} \quad (6)$$

so the equation becomes

$$(1 + \delta_z^2 + \delta_r^2)T_{ij}^{n+1} = (1 - \delta_z^2 - \delta_r^2)T_{ij}^n \tag{7}$$

where the indices for $\alpha$ were "linearized" for solvability by keeping them at $n$ instead of $n$+1. The fourth order cross-derivatives are assumed to be very small and change slowly in time relative to $\Delta t$, which should be valid in realistic cases since the heat equation is diffusive and dissipative (DuChateau and Zachmann, 2002).

$$\delta_z^2\delta_r^2 T_{ij}^{n+1} - \delta_z^2\delta_r^2 T_{ij}^n \approx 0 \tag{8}$$

Subtracting the left-hand size of Eq. (8) from Eq. (7) and collecting terms,

$$(1 + \delta_z^2 + \delta_r^2)T_{ij}^{n+1} = (1 - \delta_z^2 - \delta_r^2)T_{ij}^n - \left(\delta_z^2\delta_r^2 T_{ij}^{n+1} - \delta_z^2\delta_r^2 T_{ij}^n\right)$$

$$(1 + \delta_z^2 + \delta_r^2 + \delta_z^2\delta_r^2)T_{ij}^{n+1} = (1 - \delta_z^2 - \delta_r^2 + \delta_z^2\delta_r^2)T_{ij}^n$$

$$(1 + \delta_z^2)(1 + \delta_r^2)T_{ij}^{n+1} = (1 - \delta_z^2)(1 - \delta_r^2)T_{ij}^n \tag{9}$$

$T_{ij}^*$ is defined apart from the constants of integration by the relationship,

$$(1 + \delta_z^2)T_{ij}^* = (1 - \delta_r^2)T_{ij}^n \tag{10}$$

which is substituted into the right-hand side of (9),

$$(1 + \delta_z^2)(1 + \delta_r^2)T_{ij}^{n+1} = (1 - \delta_z^2)(1 + \delta_z^2)T_{ij}^* \tag{11}$$

The derivatives commute,

$$(1 + \delta_z^2)(1 + \delta_r^2)T_{ij}^{n+1} = (1 + \delta_z^2)(1 - \delta_z^2)T_{ij}^* \tag{12}$$

$(1 + \delta_r^2)T_{ij}^{n+1}$ and $(1 - \delta_z^2)T_{ij}^*$ must therefore be equal with the correct choice of constants of integration for $T_{ij}^*$,

$$(1 + \delta_r^2)T_{ij}^{n+1} = (1 - \delta_z^2)T_{ij}^* \tag{13}$$

Each term in this system of equations,

$$(1+\delta_z^2)T_{ij}^{*} = (1-\delta_r^2)T_{ij}^{n}$$
$$(1+\delta_r^2)T_{ij}^{n+1} = (1-\delta_z^2)T_{ij}^{*} \qquad (14)$$

can be represented as a tridiagonal matrix, so the tridiagonal matrix algorithm can be used to solve it efficiently.

Since this is cylindrical coordinates, the centerline *j*=0 is a special case that can be handled using the method of discretization by Scott and Ko (1968).

The model is parameterized for thermal properties of the material below. It also incorporates radiative heat transfer at its surface using albedo, emissivity, and insolation parameters following Mitchel and dePater (1994) and Vasavada et al. (1999).

## Thermal Conductivity of Regolith Without Ice

Parameterizing the model's soil properties relies upon published measurements in the literature, cited below. Those measurements show that thermal conductivity is a function of temperature, porosity (equivalently, bulk density) of the granular material, and interstitial gas pressure. Appendix A summarizes the data sets that are used in this effort.

Bulk density $\rho$ of the regolith varies on the Moon with location and depth beneath the surface, but it is not well known for asteroids. Lunar values are chosen consistent with Apollo core tubes and other Apollo measurements. Asteroid bulk densities are typically determined by fitting the results of thermal modeling to the observed thermal inertias of the asteroids. Measurements by Rosetta during flyby of 21 Lutetia indicates the thermal inertia increases below the top few centimeters "in a manner very similar to that of Earth's Moon" (Keihm et al., 2012). This could indicate particle sizing and/or bulk density variations over that depth, but apart from this very little is known of possible vertical structure in asteroid regolith. The parameters in this model can be adjusted to match future spacecraft measurements to help determine asteroid regolith structure.

An important question is whether thermal conductivity also varies with particle size distribution. Chen (2008) measured thermal conductivity in terrestrial sands with different particle size distributions, varying porosity and moisture content (only the cases with zero liquid moisture are relevant to airless bodies) at ambient pressure and temperature. The $D_{50}$ median particle size of these samples varied by a factor of about 6, and samples included some with narrower (well sorted, or uniform) and broader (poorly sorted, or well graded) distributions. The results found thermal conductivity to vary with porosity but not with particle size distribution. Presley and Christensen (1997) measured thermal conductivity in soda lime borosilicate glass beads of various sizes, varying pore gas pressure from 0.5 Torr to 100 Torr at ambient temperature. Only one porosity case was measured for each grain size, with finer particles generally forming more porous packings. Since Chen's results showed thermal conductivity is independent of particle size, the differences in thermal conductivity measured by Presley and Christensen might actually be due to the samples' porosities, not due to their grain sizes. On the other hand, Chen used realistic geomaterials while Presley and Christensen used spherical beads. It is possible that the size of contact patches between spherical particles is correlated to grain diameter, so there may be a grain size dependence that exists in Presley and Christensen's data that doesn't exist in realistic regolith. A literature review found no measurements that varied temperatures for different particle sizes while keeping constant porosities, or that varied temperatures for different porosities while keeping constant particle sizes, so for now the results by Chen are the only guidance and they indicate thermal conductivity does not vary with particle size as an independent variable for realistic geomaterials.

Thermal conductivity for actual lunar soil has been found to follow the form,

$$k = A\left(1 + \chi\left(\frac{T}{350\mathrm{K}}\right)^3\right) = A + B\left(\frac{T}{350\mathrm{K}}\right)^3 \qquad (15)$$

where $K$ is in kelvins (temperature units), and where $\chi$, $A$, and $B$ are model parameters. For example, Apollo 12 soil sample number 12001,19 was measured by Cremers and Birkebak (1971) and is shown in Fig. 1, where the dashed line is our fit using $A = 0.887$ mW/m/K and $\chi = 1.56$ ($R^2 = .9929$).

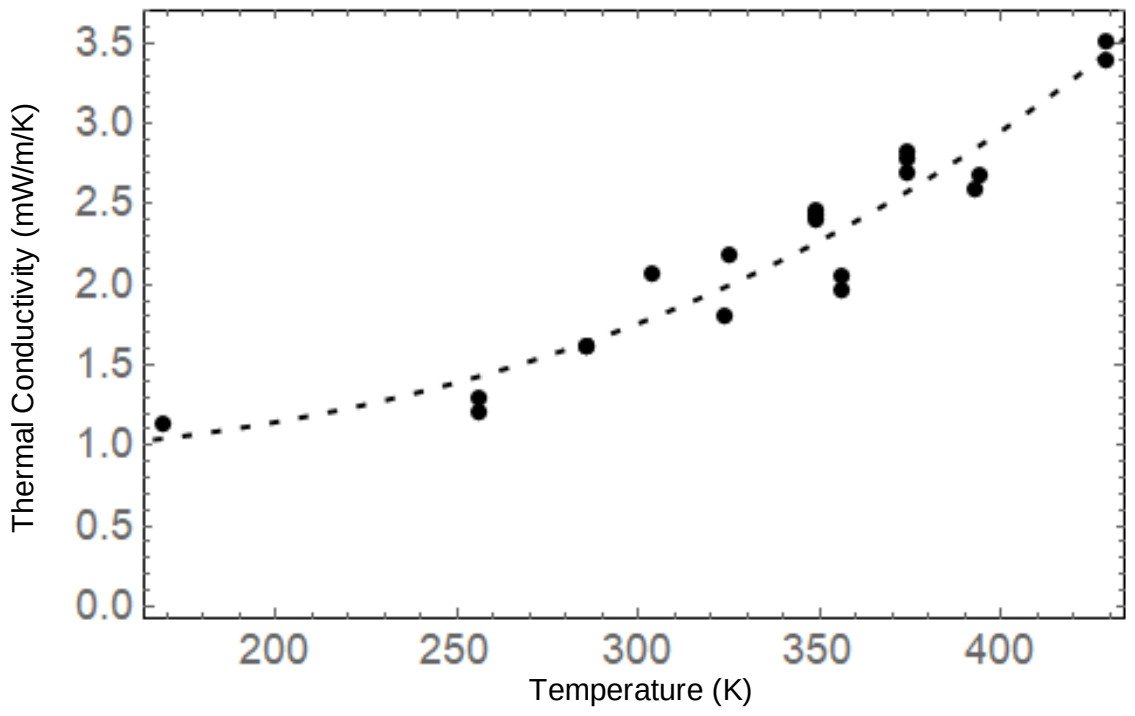

**Figure 1:** Thermal conductivity measurements of Apollo soil sample 12001,19.

Apollo 14 soil sample 14163,133 was measured by Cremers (1972) at two different bulk densities as shown in Fig. 2: 1100 kg/m$^3$ (black dots with dashed curve fit, $R^2 = .9934$) and 1300 kg/m$^3$ (open circles with gray curve fit, $R^2 = .9926$). These densities are only 17% different and considering the difficulty of maintaining local density in an experimental apparatus this appears inadequate to identify a trend.

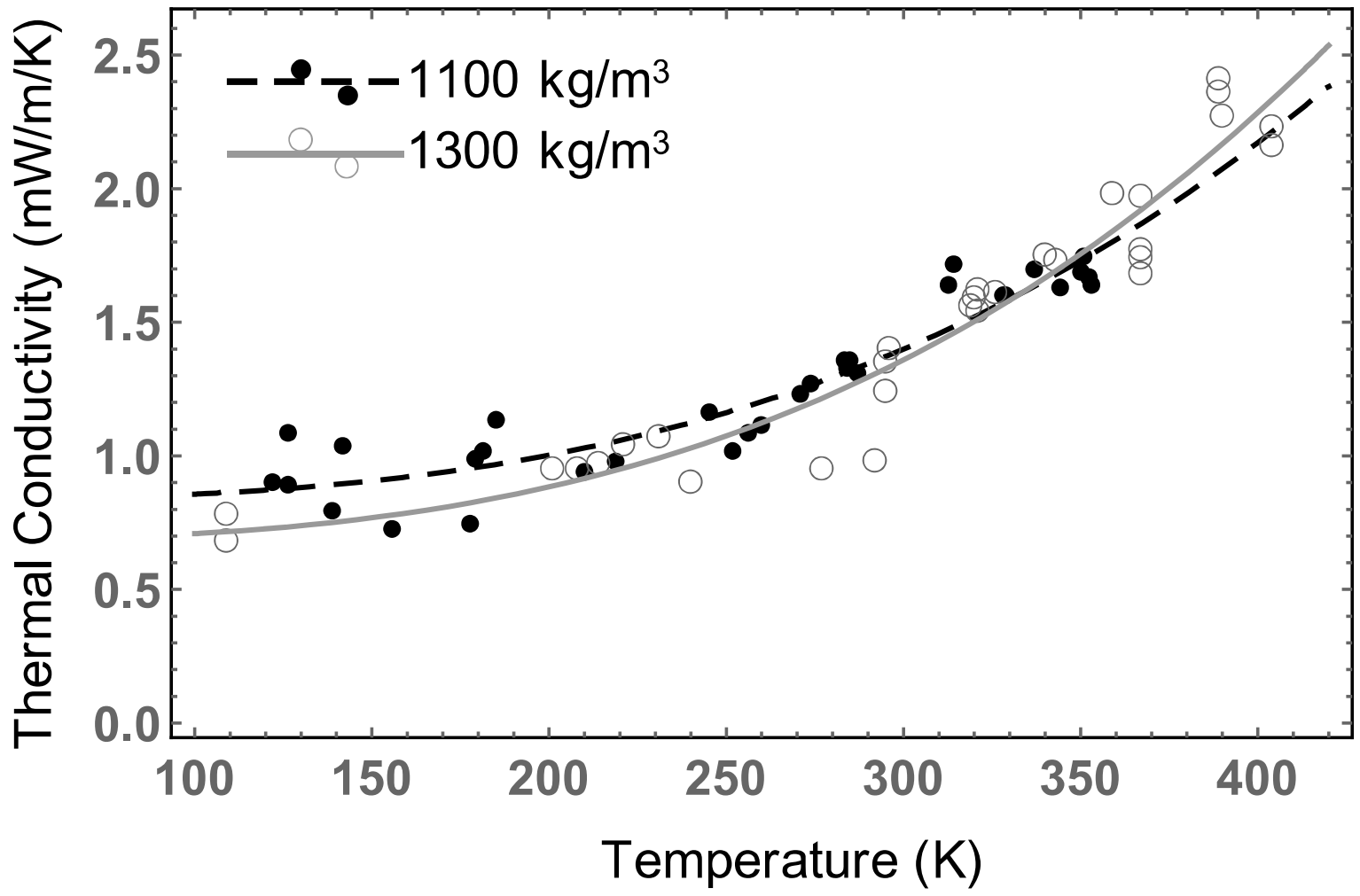

**Figure 2:** Thermal conductivity measurements for an Apollo 14 soil sample at two different bulk densities.

A greater variation of densities was measured by Fountain and West (1970) using crushed basalt in 10 torr vacuum at six different bulk densities as shown in Fig. 3: $\rho_1 = 790$ (pluses), $\rho_2 = 880$ (circle), $\rho_3 = 980$ (down-triangles), $\rho_4 = 1130$ (squares), $\rho_5 = 1300$ (diamonds) and $\rho_6 = 1600$ kg/m$^3$ (up-triangles). Note the 980 and 880 kg/m$^3$ samples do not follow the trend of decreasing thermal conductivity shown by the other samples, probably due to experimental uncertainty.

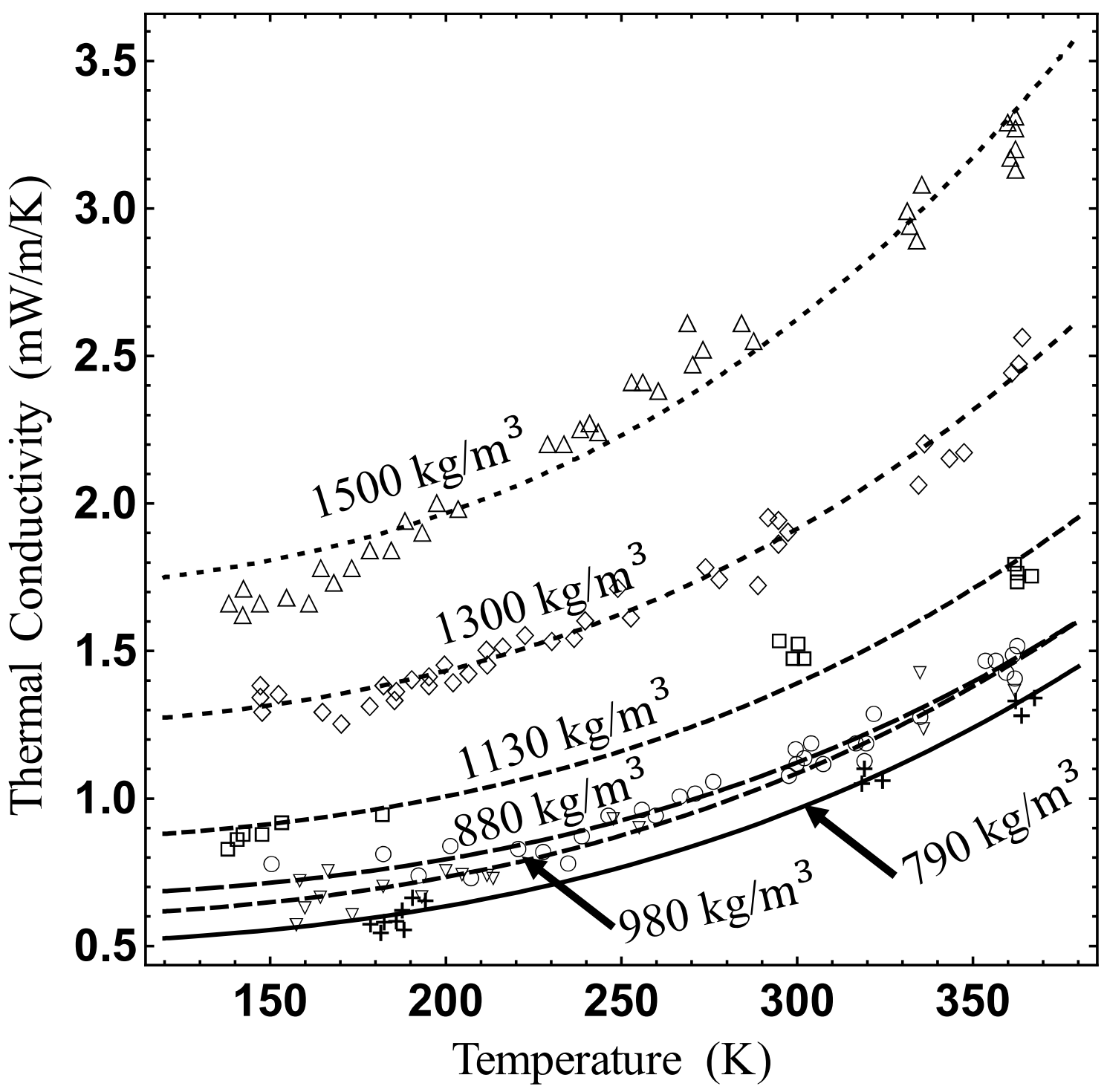

**Figure 3.** Thermal conductivity for six bulk densities of crushed basalt.

Many functional forms were analyzed to fit all these data into one overall equation. Two forms were found to provide excellent fit and they are discussed below. The first is a power law of the porosities, and the second is an exponential of the porosities.

**<u>First Functional Form</u>**

The heat flux field may be decomposed into two contributions. The first is the flux that would exist if radiative heat transfer could be switched off. A hypothesis is that parameter $A$ in Eq. 15 should scale as a power law of the solid fraction of the material,

$$A = A_0(1 - \nu)^a \qquad (16)$$

where $\nu$ is soil porosity so $(1 - \nu)$ is the solid fraction, and $A_0$ and $a$ are model parameters obtained by fitting the data. The second heat flux contribution is the additional flux field if radiation were switched back on. That additive flux includes both the radiative field in pore spaces and the additional flux in the solid material that provides continuity to the pore flux. A hypothesis is that this passage through both the solid and radiative regions produces the product of a power law of the solid fraction and a power law of the pore fraction, so the parameter $B$ from Eq. 14 scales as,

$$B = B_0(1-\nu)^a \nu^b \tag{17}$$

where $B_0$ and $b$ are model parameters obtained by fitting data, and $a$ is the same value as in Eq. 16. Thus, defining $\chi_0 = \frac{B_0}{A_0}$, the parameter $\chi$ from Eq. 14 scales as,

$$\chi = \chi_0\, \nu^b \tag{18}$$

The $A_i$ and $\chi_i$ fitting parameters for the six fitted curves $(i = 1,2,\ldots,6)$ in Fig. 3 were themselves fitted to Eqs. (17) and (18), with the result,

$$A = 7.02(1-\nu)^{2.08}\left(\frac{\frac{\text{mW}}{\text{m}}}{\text{K}}\right), \qquad \chi = 2.16\,\nu^{1.44} \tag{19}$$

with $R^2 = 0.9948$ and $R^2 = 0.9901$, respectively. Choosing integer values $a = 2$ and $b = 1$ also produces excellent fits to the data with $R^2 = 0.9946$ and $R^2 = 0.9881$, respectively, so the integers were chosen for elegance. The best fits with these exponents are shown in Fig. 4 and are

$$A_i = 6.122\,(1-\nu_i)^2 \quad \text{and} \quad \chi_i = 1.82\,\nu_i \tag{20}$$

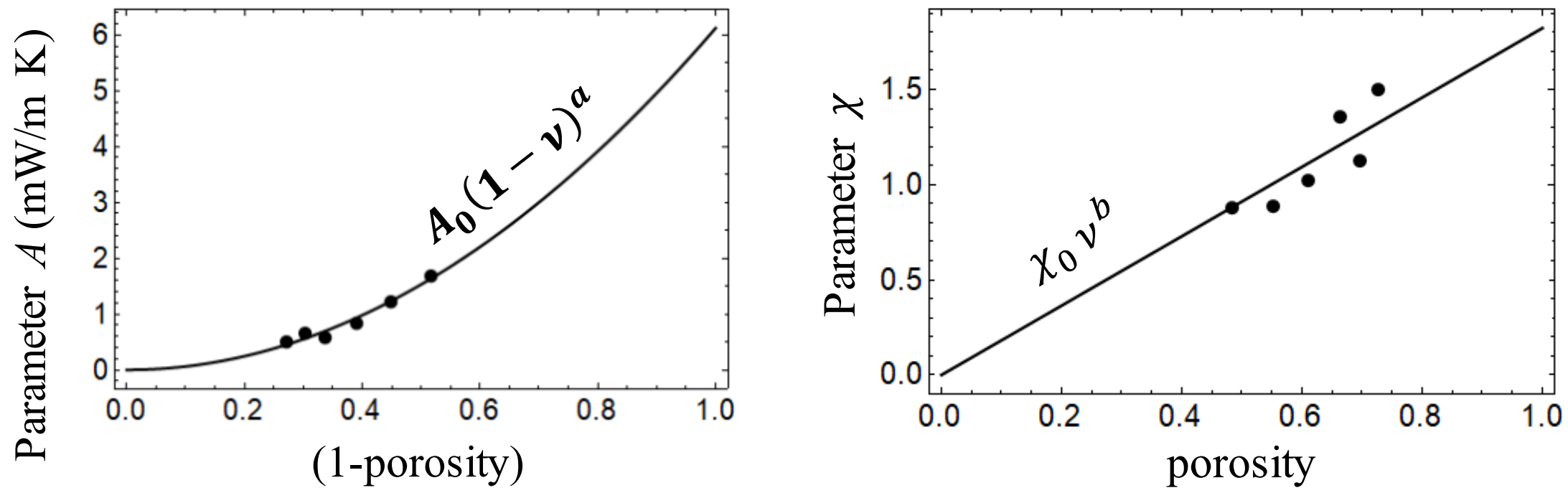

**Figure 4.** Meta-fitting of the curve-fitting: (Left) $A$ parameter, (Right) $\chi$ parameter.

The first form of the generalized function for thermal conductivity is therefore,

$$k(\nu, T) = A_0(1-\nu)^2\left[1 + \chi_0\nu\left(\frac{T}{350\text{ K}}\right)^3\right] \tag{21}$$

where $A_0 = 6.12 \times 10^{-3}$ W/m-K and $\chi_0 = 1.82$ for the basalt measured by Fountain and West (1970) but should be generally different for other materials. This equation should be valid over a range wider than the range of the data to which it was fitted, but characterizing the useful extrapolation range is beyond the scope of this work. Note also that $A_0$ cannot be interpreted as the thermal conductivity of the solid material by setting $\nu = 0$; basalt's thermal conductivity is about 400 times larger than this value. The limit $\nu \rightarrow 0$ cannot be used this way because the net contact area between grains is determined not only by $\nu$ but also by the size of asperities on the grains' surfaces.

To test Eq. (21), it is plotted in Fig. 5 against the data of Fountain and West (1970). More experimental work is needed to explain why some subsets of the data do not fit as well as others (e.g., 790 kg/m$^3$ data, or the middle temperatures of the 1500 kg/m$^3$ data). The equation is theoretically elegant but it is possibly too simple, or the experiment data might have errors due to sample handling or other deficiencies.

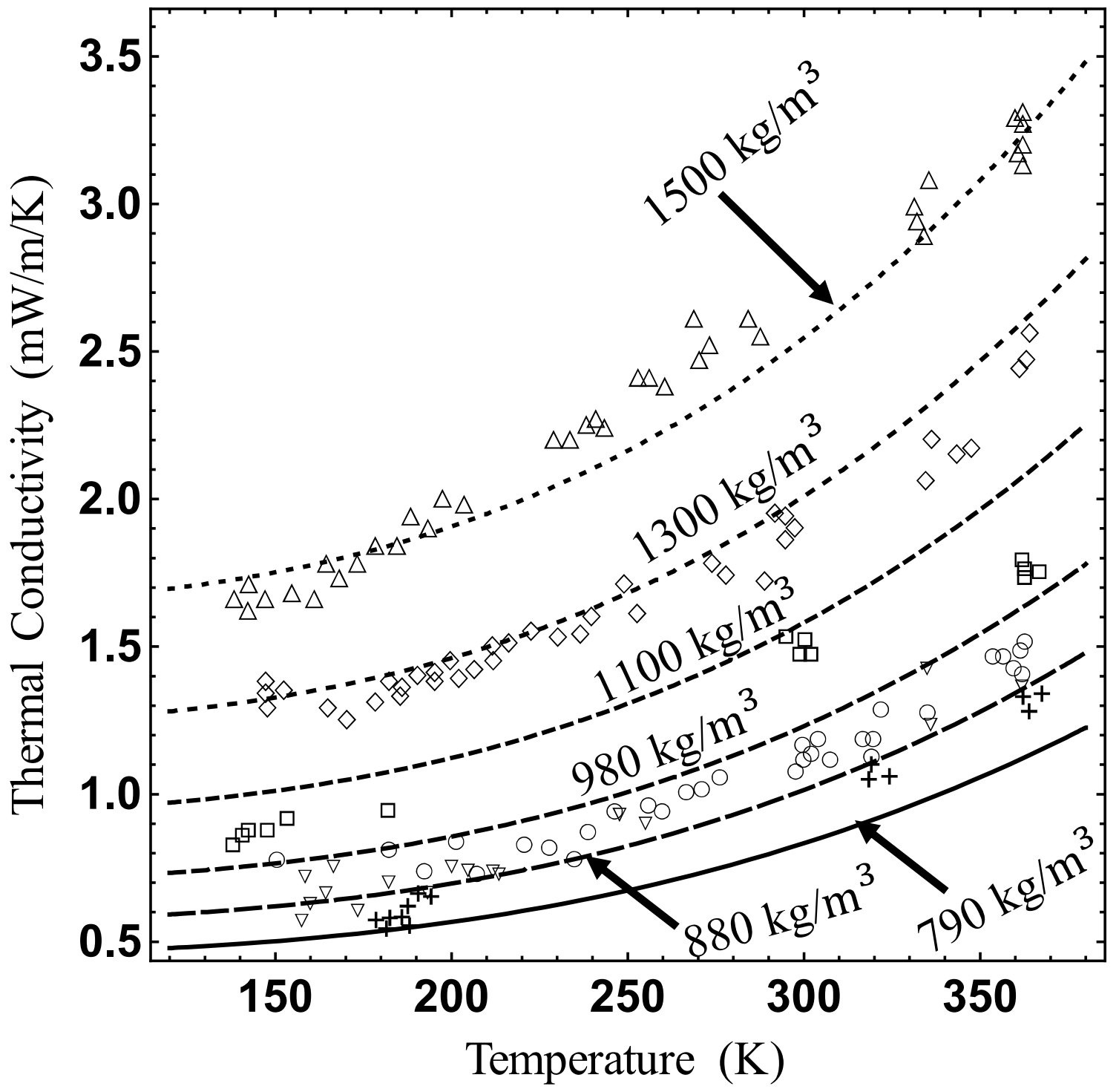

**Figure 5.** Curve fitting using power laws of porosities.

## <u>Second Functional Form</u>

The second functional form follows Chen (2008), which analyzed terrestrial soils at Earth-atmospheric pressure and temperature while varying porosity and moisture content, $S_r$,

$$k(\nu, S_r) = \hat{a}^{(1-\nu)}\hat{b}^{\nu}[(1-\hat{c})S_r + \hat{c}]^{\varepsilon\nu} \quad (22)$$

Chen found excellent fit using $\hat{a} = 7.5$, $\hat{b} = 0.61$, $\hat{c} = 0.0022$, and $\varepsilon = 0.78$. Lunar and asteroid regolith in vacuum are incompatible with liquid moisture content, so $S_r$=0, which simplifies the equation to an exponential decay,

$$k(\nu) = 7.5e^{-7.28\nu} \quad (23)$$

Note this lacks a separate temperature-dependent term as in Eqs. (15) and (21) because Chen's data were all at ambient temperature ($T \cong 300\text{K}$). Including this term,

$$k(\nu, T) = A\left(1 + \chi\left(\frac{T}{350\text{K}}\right)^3\right) = e^{a+b(1-\nu)}\left(1 + e^{c+d\nu}\left(\frac{T}{350\text{K}}\right)^3\right) \quad (24)$$

This fits the Fountain and West data as shown in Fig. 6 with

$$A = e^{-2.118+5.116(1-\nu)}\ (\text{mW/m/K}),\ \ \chi = e^{-1.301+2.256\,\nu} \quad (25)$$

with $R^2 = 0.9968$ and $R^2 = 0.9907$, respectfully. Simplifying,

$$k(\nu, T) = 20.036\, e^{-5.116\,\nu}\left(1 + 0.2723\, e^{+2.256\,\nu}\left(\frac{T}{350\text{K}}\right)^3\right)\ (\text{mW/m/K}) \quad (26)$$

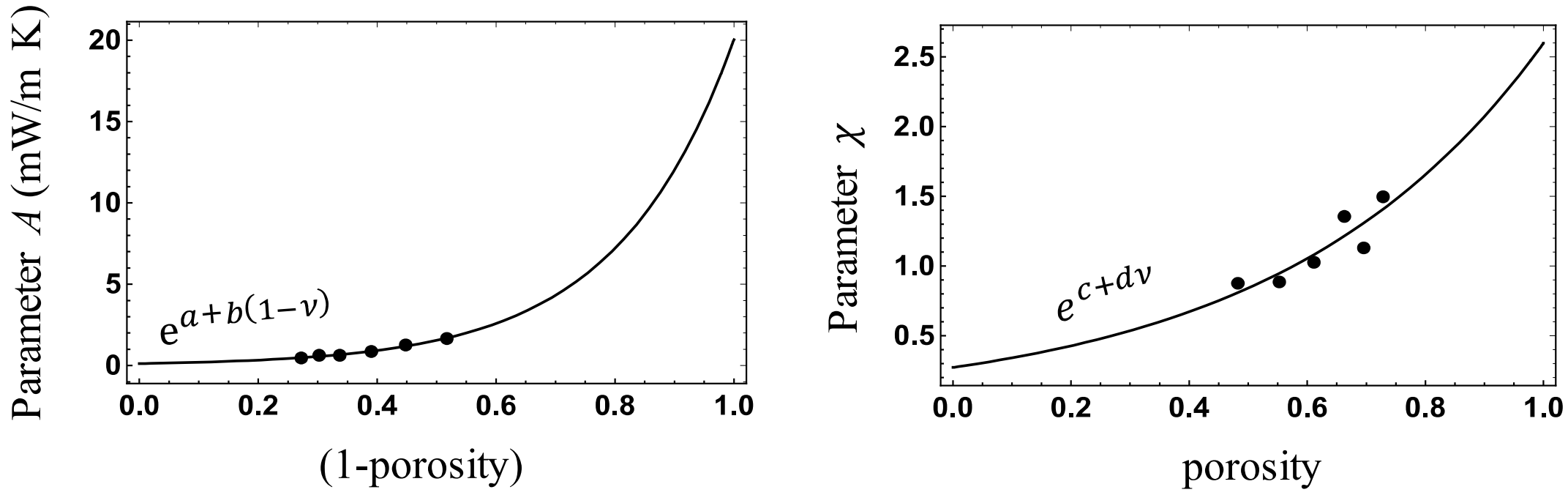

**Figure 6.** Meta-fitting for curve-fitting, where the solid lines are Eq. 24: (Left) $A$ parameter; (Right) $\chi$ parameter.

Eq. (26) is plotted in Fig. 7 against the data of Fountain and West (1970).
This provides a slightly better fit to the experimental data, but new measurements with a wider range of porosities should be diagnostic. No function that fits the data better than this has been identified, although many other forms and possible relationships were explored including proportional, linear, quadratic, and logarithmic functions of porosity, and products of power laws of porosity with power laws of solid fraction. It is possible that the data do not fit even better than this because of experimental uncertainty. It is extremely difficult to maintain constant compaction of a granular material while evacuating the pore pressure because pressure gradients can exceed the overlying weight of soil both macroscopically and microscopically (locally). Also, thermal conductivity measurements can change the compaction of soil because thermal cycling causes grains to expand and contract, and prior work with granular materials (Chen et al., 2006) and lunar soil simulants (Gamsky and Metzger, 2010, Metzger et al., 2018) shows this is an effective compaction mechanism. Metzger et al. (2018) found it extremely difficult to maintain low compaction states of lunar simulant because even tiny mechanical shocks cause internal avalanches and compaction. Therefore, it may be difficult to obtain experimental results that fit better than in Fig. 7. The Apollo lunar soil data in Fig. 2 were checked and they are well-fitted by Eq. (26). For now, this second form is selected for the remainder of the study.

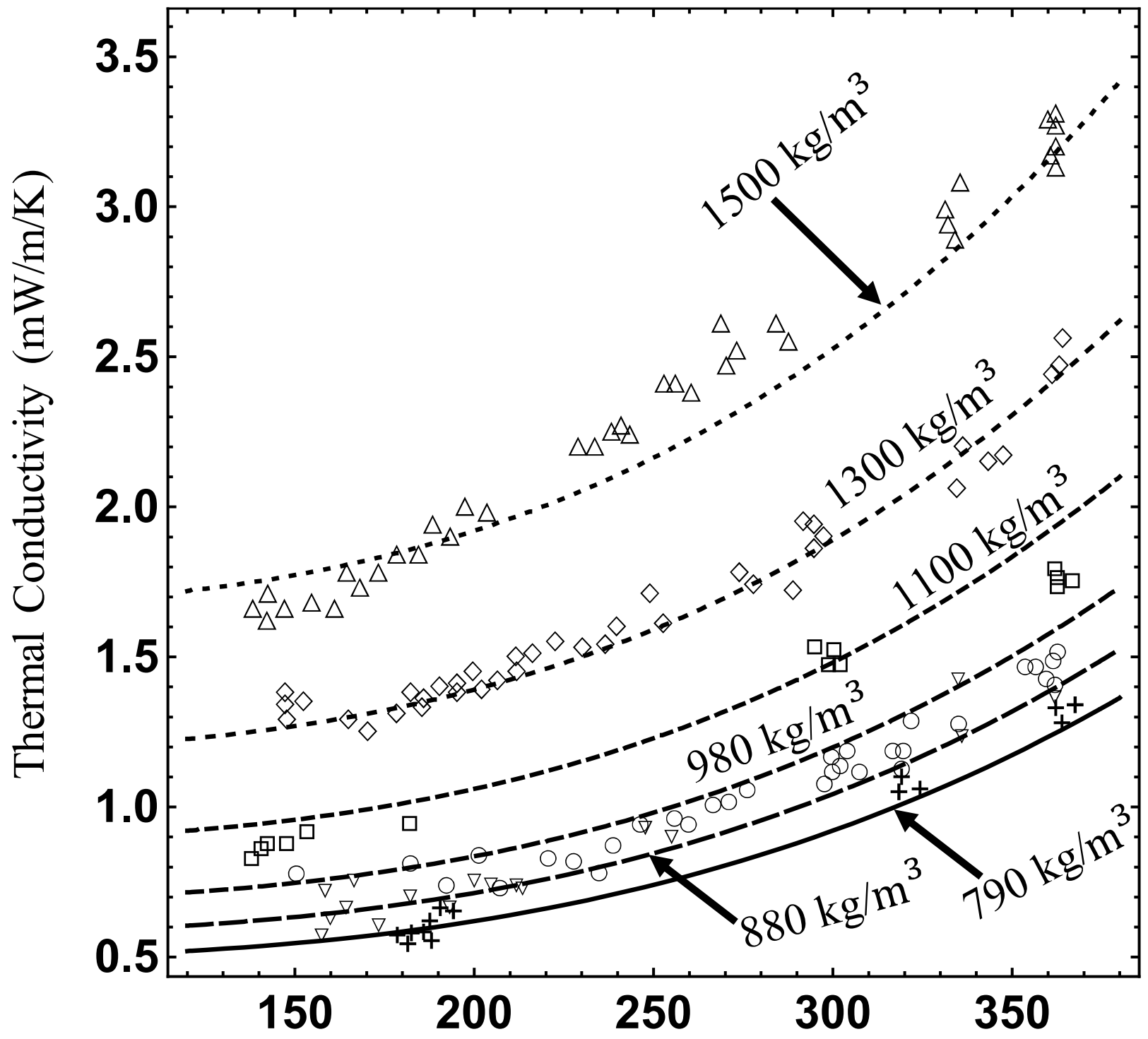

**Figure 7.** Curve fitting using exponentials of porosities.

## Comparison with Other Data Sets

Fig. 8 compares data from Fountain and West (1970) (FW), Presley and Christensen (1997) (PC), and Chen (2008). The pore pressure differences are discussed in the section below on pore pressure dependence. This section discusses discrepancies in the forms of the curves. The top black points are from Chen (2008) data for 4 sand samples of varied particle size in four packing porosities, each, with no moisture content, at ambient pressure (~760 Torr), and at ambient temperature (~300 K). The solid curve is the fit by Chen evaluated for no moisture content, and the curve is dashed where extrapolating beyond the measurements. The middle graphs (from 100 Torr to 0.5 Torr) are from PC with glass spheres in 8 samples each having a different mean particle diameter (each sample is a vertically-aligned set of points correlated to one porosity value) measured at 17 pore pressures (the lines connect different samples at the same pore pressure as a guide to the eye) and ambient temperature (~300 K). The bottom black solid line is Eq. (26) fit to FW at $10^{-8}$ Torr, evaluated here at $T$=300 K for consistency with Chen and PC, dashed where extrapolating beyond the range of measured porosities. The error bars were calculated for the 6 porosities where FW measurements were taken.

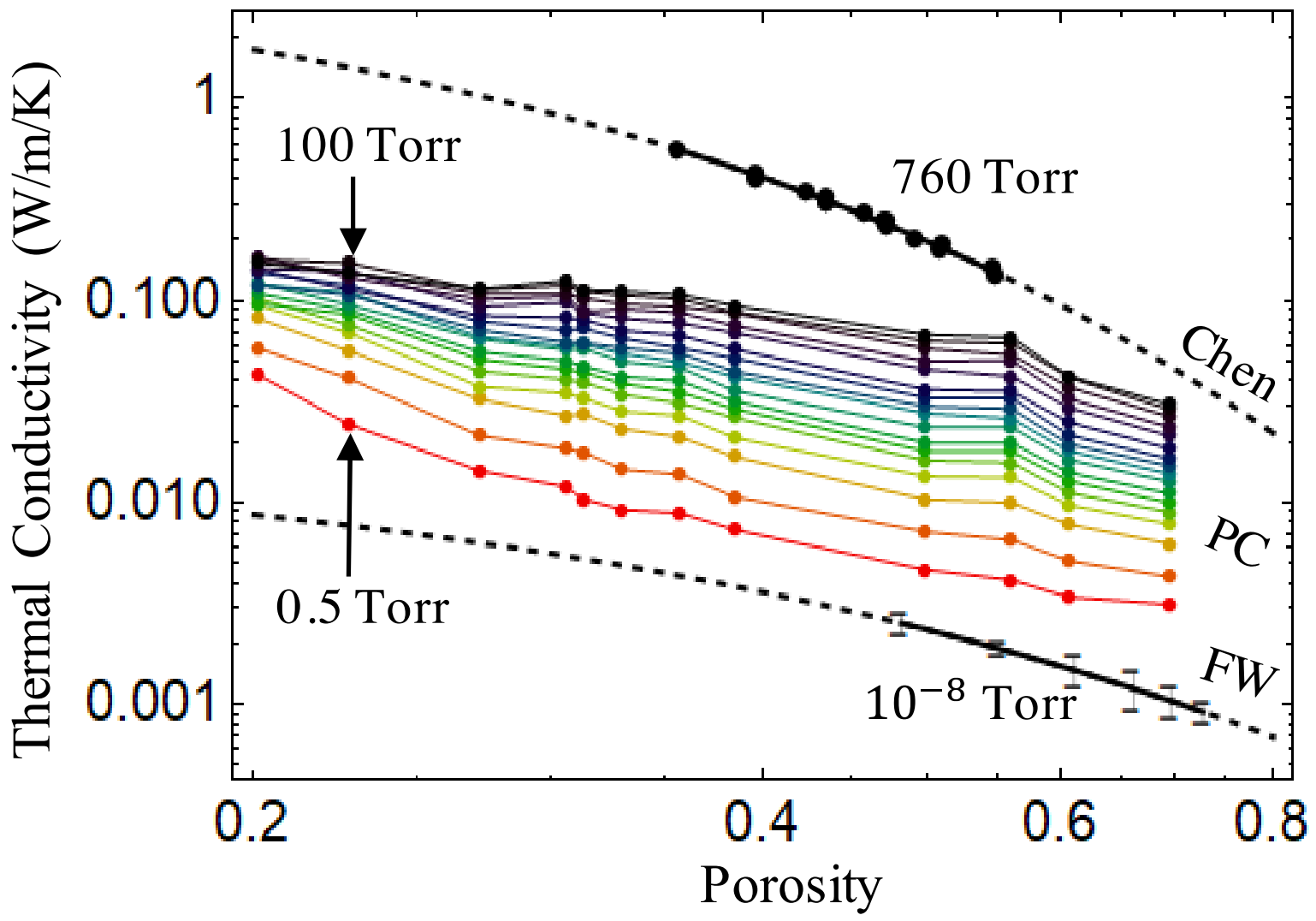

**Figure 8.** Thermal conductivity vs. porosity at different pore pressures, comparing the three data sets.

PC does not fit the form of Eq. (26), but Chen and FW both fit that form. This might be explained by the difference in particle shapes. PC samples were smooth spheres, which under compression have large contact patches that are a function of particle diameter, whereas the Chen and FW samples were geologic materials with irregular shapes and asperities that create much smaller contact patches uncorrelated to particle diameter. However, larger contact patches should produce greater $k$, but in PC they produced smaller $k$ than the trends of Chen and FW.

Alternatively, the fact that PC does not fit the form of Eq. (26) might be explained as experimental disturbances affecting the coarse particles in PC more than the fine particles. This could happen either through random mechanical vibrations in the laboratory environment (Metzger et al., 2018) or by the drag forces of gas permeation since PC measurements were taken at a variety of pressures. The Kozeny-Carman relationship [Kozeny, 1927; Carman, 1956; Carrier, 2003] applied to the PC data shows permeability would be two orders of magnitude greater for the coarsest (but least porous particles) than for the finest (but more porous) ones, so gas drag forces would be two orders of magnitude weaker for the larger particles. Cohesive energy to stabilize a granular packing scales as the number of contacts per volume, which scales as the inverse of particle diameter cubed and decreases with porosity due to the decreasing grain contact coordination number by a factor of three over the range of porosities in PC (Murphy, 1982). Cohesive energy per grain contact scales proportionally to particle diameter for spheres (Walton, 2007). Overall, cohesive forces scale as two and a half orders of magnitude stronger for the finest particles than for the coarsest ones. This rough analysis indicates the finest PC particles should be more resistant to gas permeation disturbance by a factor of five compared to the coarsest particles, and more resistant to incidental mechanical shock and vibration disturbance by a factor of 240 compared to the coarsest particles. This supports the hypothesis that the coarser particles (lower porosities in Fig. 8) were more disturbed during the experiments, affecting the shapes of the curves. The reduced thermal conductivity for the coarser particles (relative to the trend lines of Chen and FW) suggest these cases were more porous than believed, indicating the gas exiting the material during vacuum pump-down fluffed these cases and reduced their grain-to-grain contacts.

The consistent curve shape for FW and Chen provides confidence that Eq. (26) can be extrapolated modestly beyond the range of porosities measured by FW. The combined range of porosities measured by Chen and FW is $0.355 < \nu < 0.73$. Lunar soil bulk densities are primarily in the range $900 < \rho < 2200$ kg/m$^3$, corresponding to $0.26 < \nu < 0.71$, so only modest extrapolation is required at the low end of the range where Chen data provide high confidence in the functional form. However, a thin surface veneer of *epiregolith* may exist globally on the Moon, a “fairy castle” state with $\nu \sim 0.9$ made possible by low gravity and photoionization in the strong ultraviolet light (Mendell and Noble, 2010). Also, experimental work suggests surficial regolith may be more porous in PSRs due to the absence of thermal cycling (Gamsky and Metzger, 2010; Metzger et al., 2018). The impact dynamics of the LCROSS spacecraft in Cabeus crater, a PSR, suggests $\nu \sim 0.7$ to a depth of two or more meters. If the geologic processes of a PSR compacted it to only $\nu \sim 0.7$ with such overburden, it is possible the soil may be even less compacted in the upper layers where there is less overburden. Extrapolation $\nu > 0.73$ via $k(\nu, T)$ may therefore be needed, and further laboratory measurements should be performed to validate the model over the wider range of porosities.

## Thermal Conductivity of Lunar Ice

For asteroids, the volatile molecules are bound in the crystalline structure of the hydrated minerals and not generally in the form of physical ice. For the Moon, the volatiles include adsorbed molecules on the surfaces of the grains as well as solid ice mixed in the regolith. The contribution of ice to thermophysical properties of regolith is determined by its chemistry and its physical state: amorphous or crystalline, "snow" mixed in the pore spaces, solid ice cobbles like hail, etc. The thermophysical properties of amorphous ice can vary by orders of magnitude depending on density and microstructure (Mastrapa et al., 2013). Amorphous ice crystallizes exothermically when there is adequate activation energy. This has been considered a mechanism for comet outbursts as heat diffuses into the interior reaching amorphous material (Sekanina, 2009). In the lunar case, impact gardening could provide the activation energy crystallizing the deposits as it matures. For now, this thermal model will be based on the geological picture presented by Hurley et al. (2012), that the ice began as a homogeneous sheet and was fragmented by impact gardening, mixing grains of pure crystalline ice among grains of otherwise dry soil. The LCROSS impact did detect crystalline ice in the ejecta (Anthony Colaprete, personal communication, 2016). If the fragments are smaller than a volume element in the model, then a volumetric mixing model is adequate. These mixing models have been investigated for icy regolith by Siegler et al. (2012).

The composition of lunar ice was calculated by Tony Colaprete (personal communication, 2016) of NASA on the basis of LCROSS impact ejecta measurements by combining measurements from the two instruments (Colaprete et al., 2010; Gladstone et al., 2010). The calculated volatile concentrations are shown in the Table 1. Hydrogen gas was detected, but it should not be stable even at the temperatures of the lunar polar craters. The hydrogen gas and hydroxyl may have been products of chemistry driven by heat of the LCROSS spacecraft impact. It is beyond our present scope to back-calculate what chemicals must have been present in the ice prior to the impact. For now, this remains the best estimate of the composition of lunar ice.

The saturation curves of these volatiles (NIST, 2017) shown in Fig. 9 illustrate that temperatures adequate to release water from the regolith will also release many other volatiles. For now, only water sublimation has been modeled. Modeling here may treat the sublimation of each species separately based on partial pressures and treat the diffusion of gas through the pore spaces based on overall pressure and molecular collision rates in the mixed gas. This assumption needs to be checked with measurements of actual lunar ice. Kouchi et al. (2016) found that ice mixtures of CO to $H_2O$ in ratios 50:1 and 10:1, subjected to conditions for sublimation of the CO but not the $H_2O$ left the water ice in a porous amorphous state with density similar to high-density amorphous ice. They also found that at 140K this matrix-sublimated high-density amorphous ice transitioned to cubic ice. Doubtless, the porosity resulting from matrix sublimation will decrease thermal conductivity of the remaining matrix, and transitioning back to crystalline ice will increase it, so these effects may occur when subliming mixed composition lunar ice. The non-water species of lunar ice constitute less than 50%wt of the combined ices, so the induced porosity should be much less than reported by Kouchi et al. For now, the effect is ignored. Future work may add an ad hoc parameter to treat it simplistically, but it would be little more than a guess. To inform a better model, experimental measurement is needed for thermal conductivity of matrix sublimed, mixed composition ice.

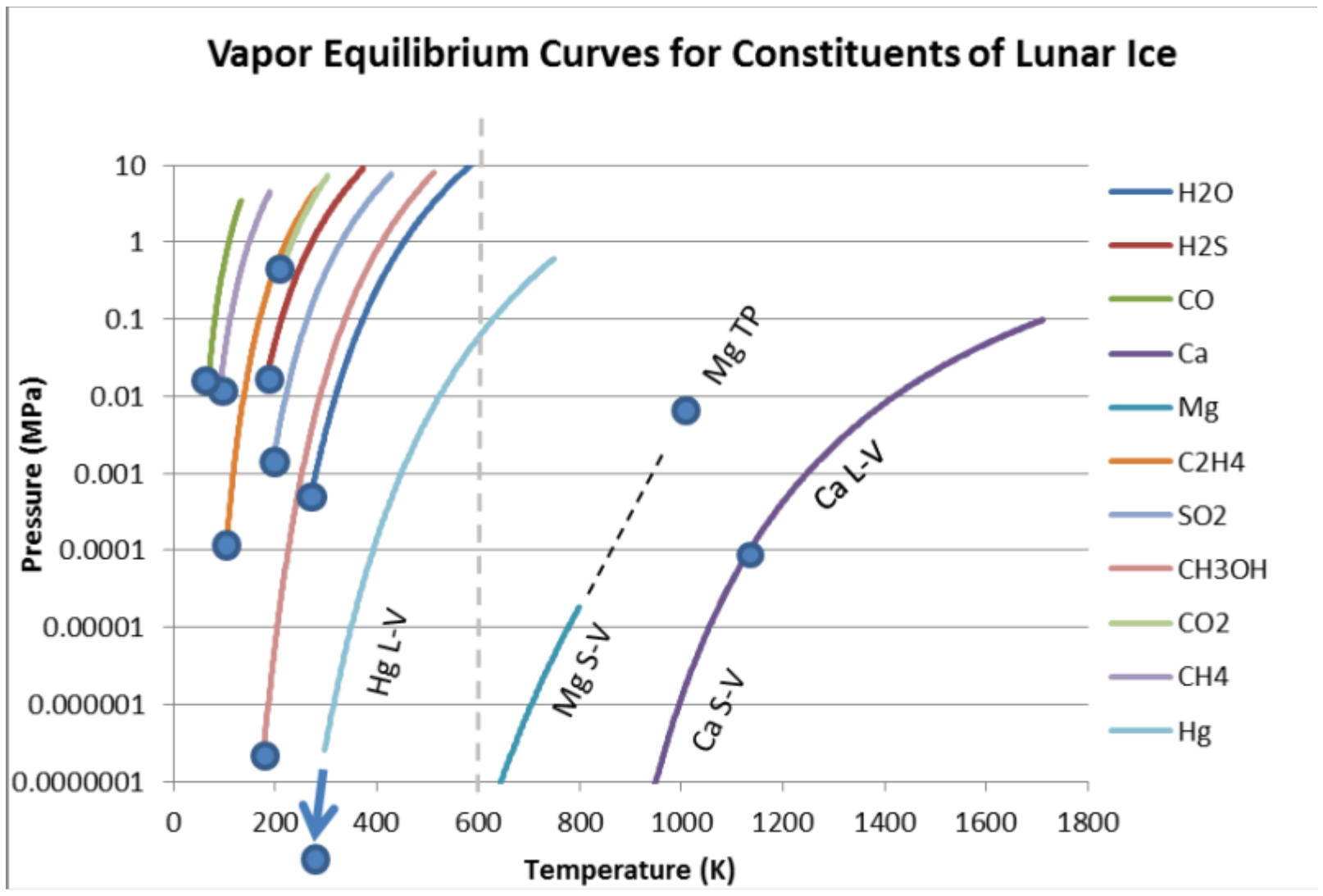

**Figure 9.** Saturation curves for chemicals in the lunar ice. (L-V = liquid-vapor, S-V = solid-vapor, TP = Triple Point, blue dots).

For thermal conductivity the contribution of pure crystalline water ice is calculated using the data points of Ehrlich et al. (2015), reproduced in Fig. 10 with the added curve fit,

$$k = 1.582 + 11.458 \exp\left(-\frac{T}{95.271}\right) \tag{27}$$

in W/m/K, and where temperature $T$ is kelvins. Available thermal conductivity data for the other volatiles are limited. Sumarakov et al. (2009) measured CO ice in the range 1K to 20K. At 20K it is about 0.5 W/m/K, more than an order of magnitude less than the extrapolation of water to that temperature per Fig. 10. Koloskova et al. (1974), cited in Sumarokov et al. (2003), measured $CO_2$ and found it about 1 W/m/K at 100 K, about 1/6 the value of water. Manzhelii et al. (1972) reports solid ammonia about 1.8 W/m/K at 100 K, about 1/3 the value of water. Lorenz and Shandera (2001) found ammonia-rich (~10-30%) water ice has thermal conductivity about 1/2 to 1/3 that of pure water ice.

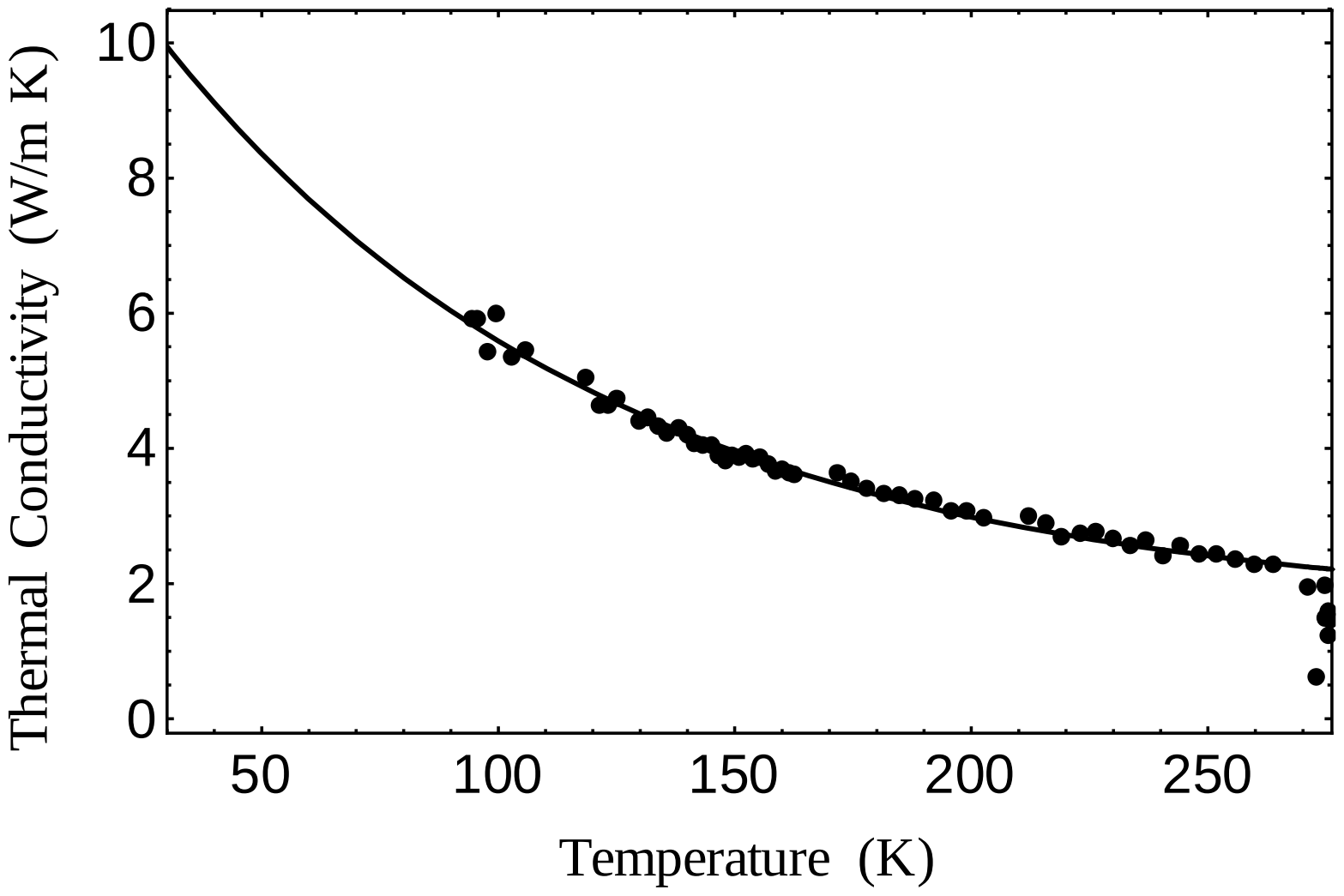

**Figure 10.** Thermal conductivity of ice.

In 1D model of sandwiched materials, the net thermal conductivity is

$$k_{\text{eff}} = \left(\sum \frac{d_i}{k_i}\right)^{-1} \tag{28}$$

where $d_i$ and $k_i$ are the thickness and thermal conductivity of each layer. To first-order approximation, which is the limit of accuracy considering the other unknowns, a mixture of dry regolith with ice grains would scale as

$$k_{\text{bulk}} = \left(\frac{V_{\text{dry regolith}}}{k_{\text{dry regolith}}} + \frac{V_{\text{ice grains}}}{k_{\text{ice grains}}}\right)^{-1} \sim \left(\frac{77\%}{O(10^{-3})} + \frac{23\%}{O(6.0)}\right)^{-1} \tag{29}$$

in W/m/K, where ~23% volume fraction of ice is derived from ~8.9%wt of ice, a rough estimate based on Table 1 approximating the mixed chemistry as if it were all water. This indicates $k_{\text{bulk}} \sim 0.0013$, only about 30% higher than dry regolith. If $k_{\text{ice grains}} = 2$ W/m/K to reflect the mixed chemistry instead of 6 W/m/K for pure water ice, then $k_{\text{bulk}} \sim 0.0013$, not measurably changed. The mixed chemistry of ice cobbles can safely be ignored for modeling thermal conductivity.

## Thermal Conductivity with Gas in the Pores

As volatiles are released, the increasing pore pressure will increase thermal conductivity by orders of magnitude as shown in Fig. 8, where FW is in hard vacuum, PC is in a range of pore pressures, and Chen is at Earth ambient pressure. The order-of-magnitude of $k$ compares reasonably for all data sets when pore pressure is accounted for, although the shape of PC disagrees with the other two as discussed above. There are inadequate data to be sure how reconcile this, but the following observations lead to a hypothesis. First, as shown in Fig. 8, the PC data with higher porosity are better distributed between the end points formed by the Chen and FW curves than they are at the lower porosities. Second, as shown in Fig. 11, the data at high porosity are continuous with the upper pressure end-point, while the low porosity data are discontinuous. Fig. 11 plots the ratio,

$$\text{Ratio} = \frac{k_{\text{PC}}(\nu,P) - k(\nu,300\text{K})}{k_{\text{PC}}(\nu,0.5\text{ Pa}) - k(\nu,300\text{K})} \tag{30}$$

where $k_{\text{PC}}(\nu, P)$ represent the PC data and $k(\nu, 300\text{K})$ is Eq. 26 based on FW data evaluated at $T$=300K (the temperature chosen to match the temperature of the PC data). $k_{\text{PC}}$ is evaluated at $P = 0.5$ Pa in the denominator, which is apparently below the "floor" where pore gas does not contribute significantly to thermal conduction in the soil as discussed below. Appended on the right side of each set of points is one data point

$$\text{Ratio} = \frac{k_{\text{Chen}}(\nu) - k(\nu,300\text{K})}{k_{\text{Chen}}(\nu,0.5\text{ Pa}) - k(\nu,300\text{K})} \tag{31}$$

where $k_{\text{Chen}}(\nu)$ is Eq. (22) with the fitting parameters found by Chen and zero moisture content. The continuity for the high porosity cases suggest a hypothesis that the high porosity cases are correct while the low porosity cases (coarse particles with less cohesion) suffered experimental disturbance. Third, it is understandable why the more porous cases of PC would be more accurate than the less porous cases because they are more stabilized by higher cohesion, as discussed above.

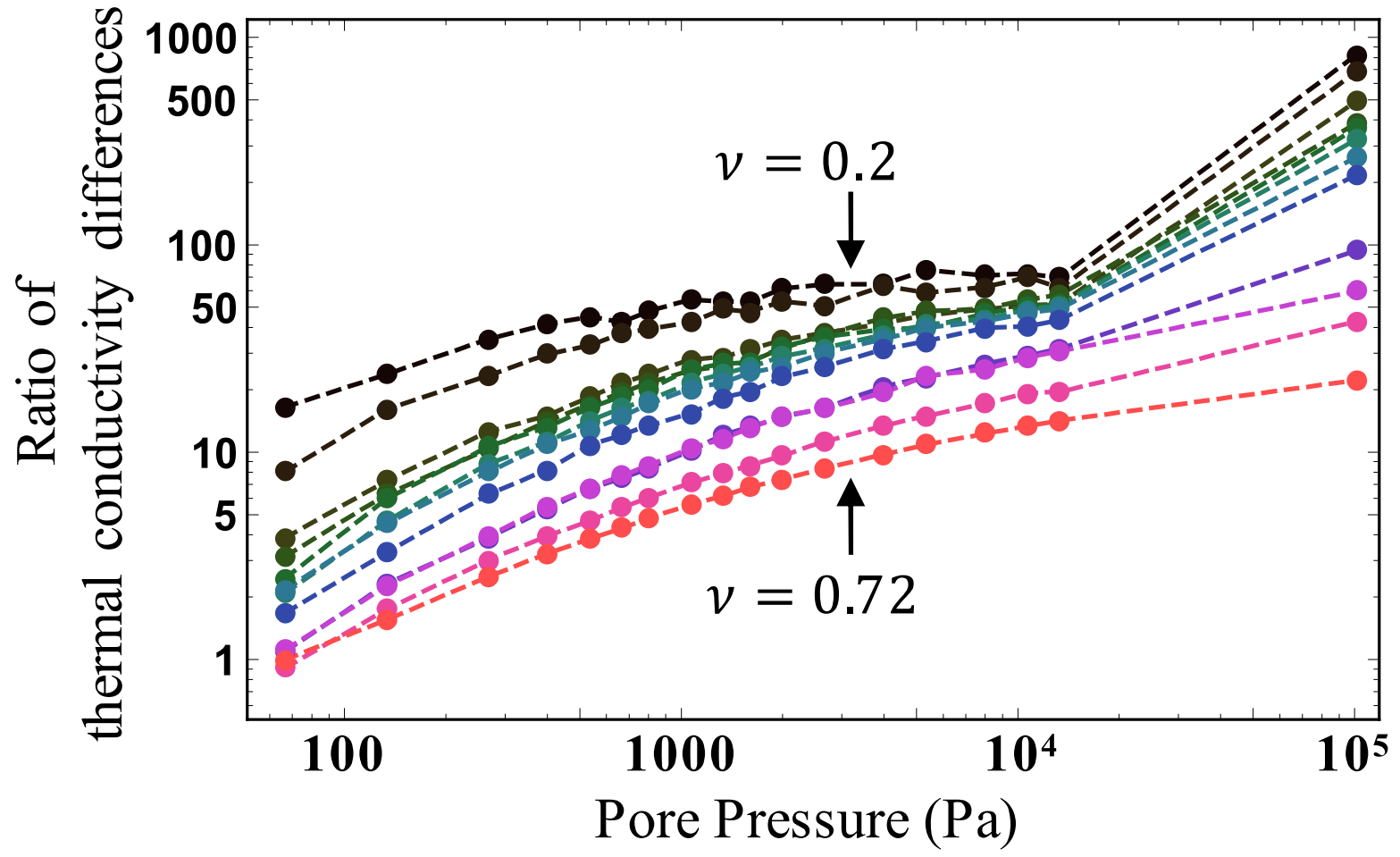

**Figure 11.** Ratio of thermal conductivity differences for two data sets.

Consistent with the assumption that the most porous data of PC are the least disturbed in the laboratory measurements, trendlines were projected on log-log axes in Fig. 12 such that they intersect at the same point where the FW and Chen trendlines are projected to intersect. These projections pass through the most porous case of PC data. Dots on the top and bottom trendlines are calculations using the Chen and FW fitted functions at the porosities of the PC data.

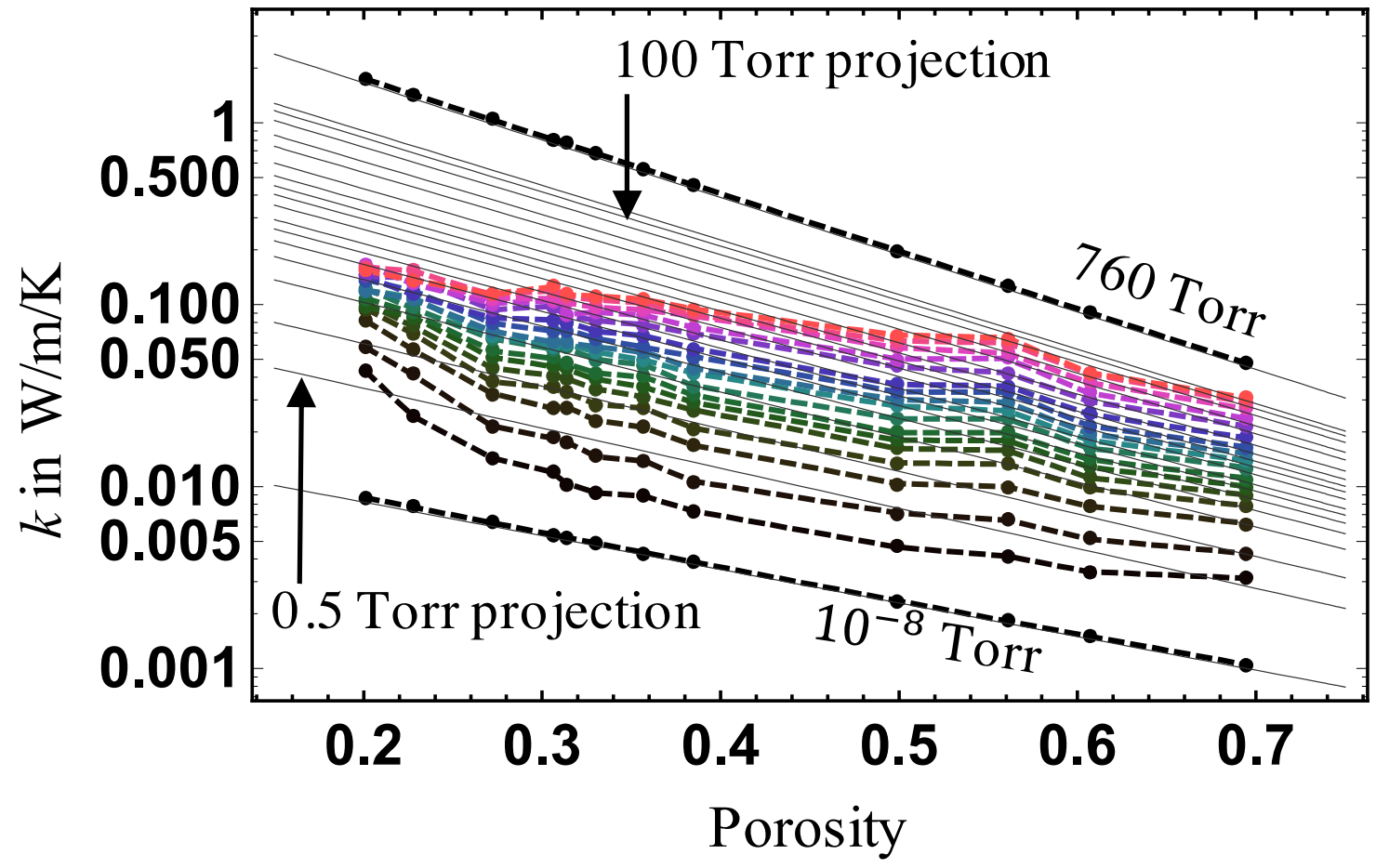

**Figure 12.** Data from Fig. 8 compared to hypothesized model (thin gray lines).

The assumption is that these trend lines are what would have been measured in PC had there been no mechanical disturbance of the samples. This assumption is necessary to reconcile the existing data sets and create a constitutive equation. The family of trend lines is viewed in Fig. 13 through two different projections: into the $(\nu, k)$ plane and into the $(P, k)$ plane, where $P$ is pore pressure in the soil. In the $(\nu, k)$ plane, the top line is for 760 torr pore pressure (Chen), the bottom is for $10^{-8}$ torr (FW), and the intermediate are for 100 torr (upper) to 0.5 torr (lower) (PC). These trendlines were chosen to intercept at the same point off the right side of the figure where FW and Chen also intercept.

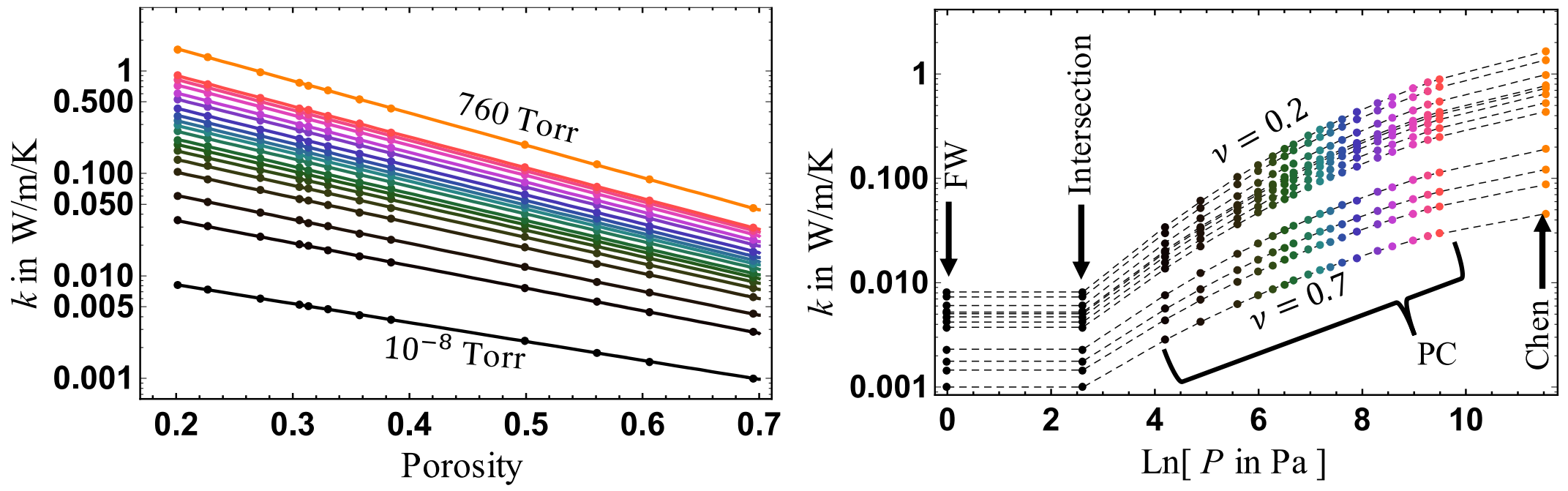

**Figure 13.** Trendlines: (Left) in the $(\nu, k)$ plane; (Right) in the $(P, k)$ plane.

In the $(P, k)$ plane the far right vertical column of points is from Chen, the left vertical column of points is from FW, and second column of points from the left is not from any dataset but is the point where the PC data project to an intersection with their corresponding “floor”. Each floor represents conduction and radiation through the grains without significant contribution from pore gas. Note that the existence of this floor implies that the gas contribution is additive to the other contributions (not multiplicative).

The general fitting function for this family of curves is,

$$k(\nu, P) = \hat{P}^{c_1}\text{Exp}\{-c_2 - c_3\,\text{Ln}^2\hat{P} + (c_4\nu - c_5)(\text{Ln}^2\hat{P} - c_6\text{Ln}\hat{P} - c_7)\} \quad (32)$$

where $\hat{P} = \text{Max}(P, P_0)$, and $P_0$ is the pressure below which is the “floor” of minimum conductivity. At constant pressure this reduces to the form of a single exponential,

$$k(\nu) = d_1 e^{d_2\nu} \quad (33)$$

whereas Eq. (26) at constant temperature reduces to the form of a double exponential,

$$k(\nu) = d_1 e^{d_2\nu} + d_3 e^{d_4\nu} \quad (34)$$

The second term is the coefficient for the radiation term in $T^3$. Radiation ought to be independent of gas pressure to good approximation in the rarefied conditions considered here so the additive form of Eq. (34) agrees with expectations. Eq. (32)

was developed from data measured at $T = 300\text{K}$. Therefore, the $T^3$ in Eq. (26) can be added to these curve-fits after subtracting the assumed $(300\text{K})^3$ contribution. With some manipulation this yields the full model,

$$k(\nu, P, T) = -k_1\, \mathrm{e}^{(-k_4\nu)}(1 - k_5 T^3) + k_6 \hat{P}^{(k_2 - k_3\nu)} \mathrm{e}^{-k_7\nu + (k_8\nu - k_9)\mathrm{Ln}^2\hat{P}} \tag{35}$$

in mW/m/K, where curve fitting with $T$ in kelvins and $P$ in pascals provided the following constants:

$$\begin{aligned} \hat{P} &= \mathrm{Max}(P, P_0) \\ P_0 &= 13.68508622330367 \text{ pascals} \\ k_1 &= 3.419683995668 \\ k_2 &= 1.3409114952195769 \\ k_3 &= 0.680957757428219 \\ k_4 &= 2.8543969429430347 \\ k_5 &= 0.000000037037037037 \\ k_6 &= 0.799089591748905 \\ k_7 &= 2.637142687697802 \\ k_8 &= 0.024344154876476995 \\ k_9 &= 0.04793741867125248 \end{aligned} \tag{36}$$

The decimal places are not all significant, but they are the exact values coded into the model. Quantifying significant digits of model parameters is left to future work when better empirical datasets are available, and when the knowledge gaps in the physics have been reduced.

## Specific Heat of Regolith Without Ice

For specific heat this model uses a mass-weighted mixing model of ice and regolith. The contribution of the dry regolith is informed by the measurements previously made for Apollo soil samples and analogue materials. Fig. 14 shows a representative comparison. Apollo samples 10084 and 10057 are from Winter and Saari (1969). The empirical fitting functions by Winter and Saari (1969),

$$C(T) = -0.034\, T^{1/2} + 0.008\, T - 0.0002\, T^{3/2} \tag{37}$$

and by Hemingway, Robie and Wilson (1973),

$$\begin{aligned} C(T) = -23.173 + 2.127\, T + 0.015009\, T^2 - 7.3699 \times 10^{-5}\, T^3 \\ +9.6552 \times 10^{-8}\, T^4 \end{aligned} \tag{38}$$

both in J/kg/K, are very close to one another and only slightly higher than the experimental data above 250 K. New fifth-order and fourth-order polynomial fits were tried. The fifth order diverges from the probable trend just outside the range of experimental measurements, so it is rejected. The fourth order is marginally better than the one by Hemingway, Robie and Wilson (HRW) and seems to preserve the trends in extrapolation. The fit by HRW is actually based on a larger set of measurements, including Apollo soil samples 14163, 15301, 60601, and 10084 to represent the average of lunar soil, so HRW is selected. It fits the data with less than 10% error.

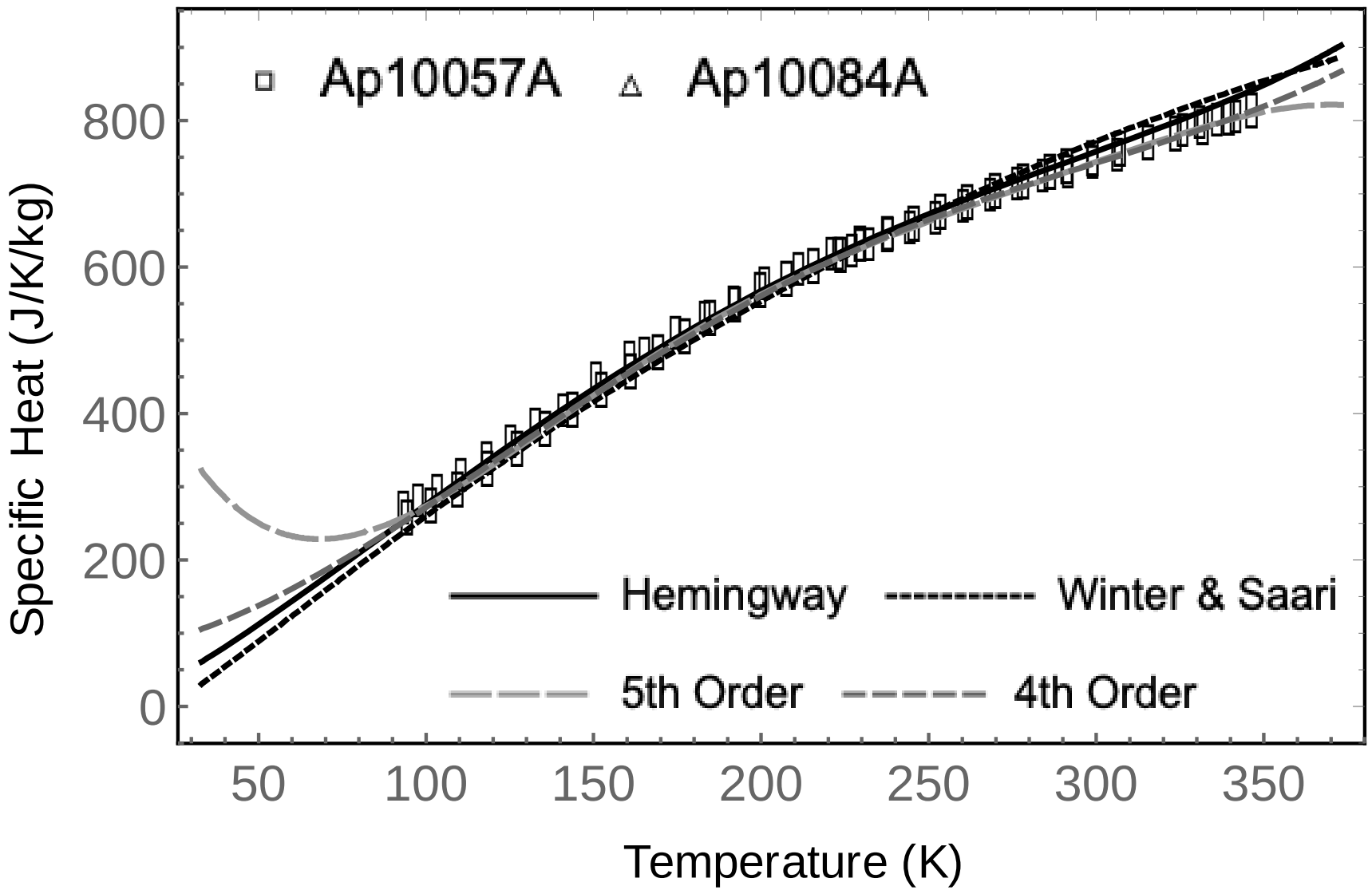

**Figure 14.** Specific heat of lunar soil vs. temperature.

The heat capacity of dry regolith should vary proportionally to the soil's solid fraction $(1-\nu)$, which varies by about ±16% from the median value in the lunar case. For the asteroid case the bulk density may vary widely due to large changes in particle size and ultra-low gravity. Measurement of asteroid regolith density in situ, including any stratigraphic variation in the subsurface, is required to guide more accurate models.

## Specific Heat of Lunar Ice

The specific heat of pure water ice was measured by Giauque and Stout (1936) from about 15K to 270K, is

$$C_{H_2O\,ice} = -100.5 + 11.43\,T + 7.101 \times 10^{-3}\,T^2 - 3.987 \times 10^{-4}\,T^3 + 2.075 \times 10^{-6}\,T^4 - 3.200 \times 10^{-9}\,T^5 \quad (39)$$

in J/kg/K, where T is in kelvins.

The specific heats of the major volatiles in lunar ice are shown in Fig. 15: (in order of prevalence) water by Giauque and Stout (1936), hydrogen sulfide by Giauque and Blue (1936), sulfur dioxide by Giauque and Stephenson (1938), ammonia by Overstreet and Giauque (1936), carbon dioxide by Giauque and Egan (1937), ethylene by Clark and Kemp (1937), methanol by Carlson and Westrum (1971), methane by Colwell et al. (1963), and carbon monoxide measured by Clayton and Giauque (1932). Weighting these according to Table 1, the composite heat capacity is shown in Fig. 16. This neglects the hydrogen and hydroxyl that were also measured in the lunar ice ejecta, which are assumed to have come from decomposition of unidentified components.

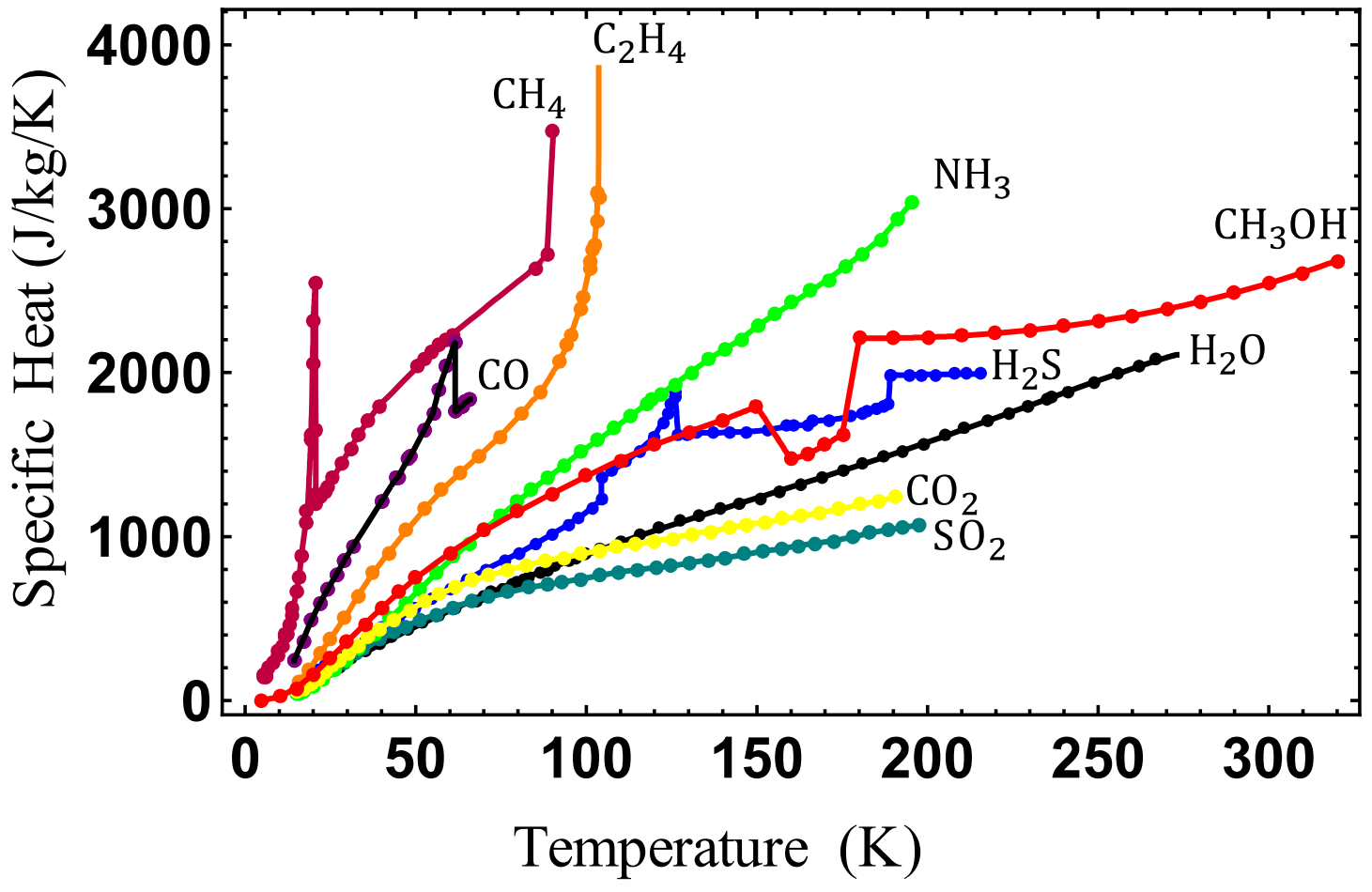

**Figure 15.** Specific heats of components of lunar ice.

**Table 1.** Volatiles in LCROSS Ejecta

| Compound | Symbol | Concentration (wt%) |
|---|---|---|
| Water | $H_2O$ | 5.50 |
| Hydrogen sulfide | $H_2S$ | 1.73 |
| Sulfur dioxide | $SiO_2$ | 0.61 |
| Ammonia | $NH_3$ | 0.32 |
| Carbon dioxide | $CO_2$ | 0.29 |
| Ethylene | $C_2H_4$ | 0.27 |
| Methanol | $CH_3OH$ | 0.15 |
| Methane | $CH_4$ | 0.03 |
| Hydroxyl | OH | 0.0017 |
| Carbon monoxide | CO | 0.000003 |
| Calcium | Ca | 0.0000008 |
| Hydrogen gas | $H_2$ | 0.0000007 |
| Mercury | Hg | 0.0000006 |
| Magnesium | Mg | 0.0000002 |

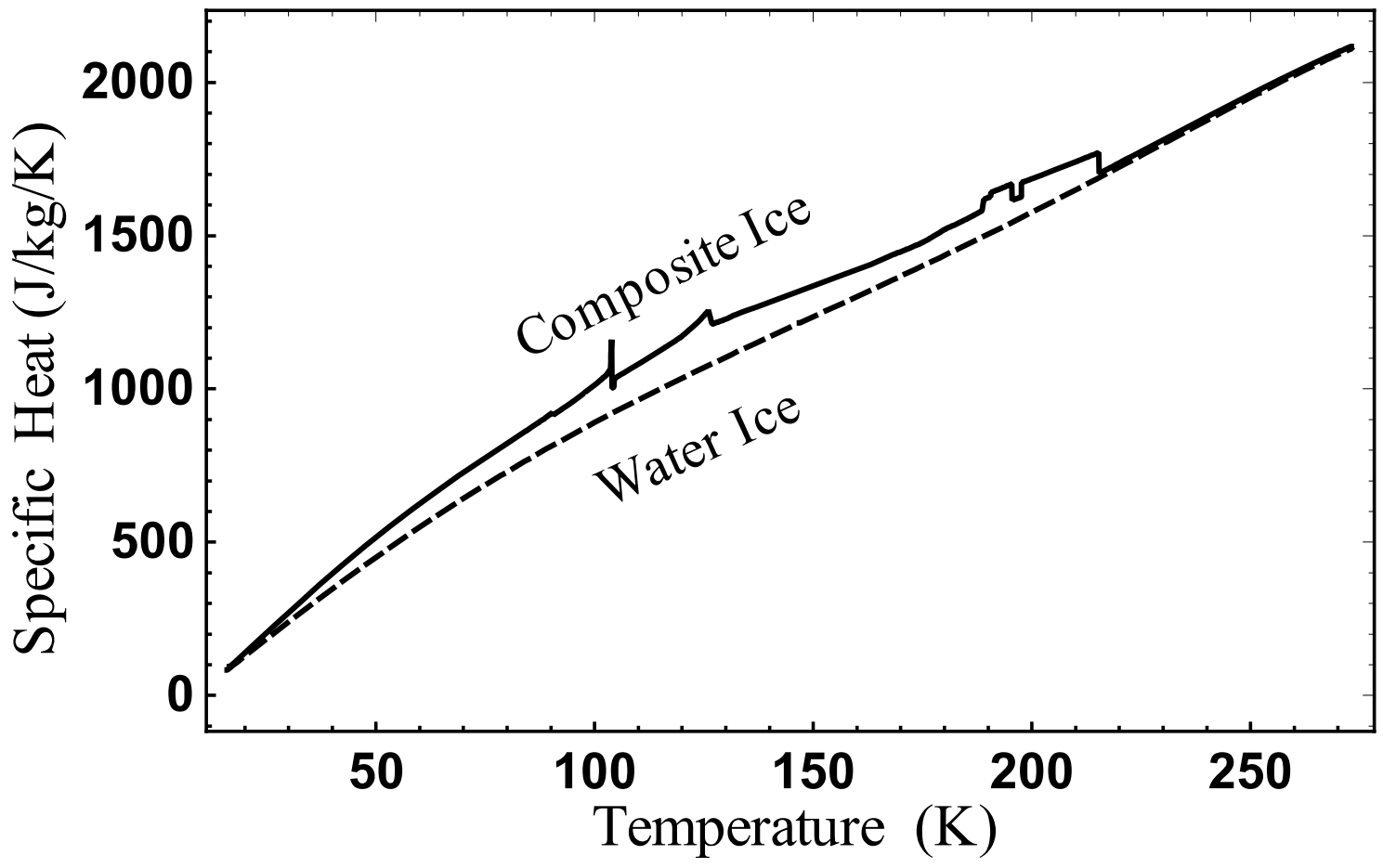

**Figure 16.** Specific heat of water ice and composite lunar ice.

The composite heat capacity for ice is calculated by a mass-weighted sum of the individual components. This assumes a linear mixing model which may not be correct depending on the actual crystalline or amorphous form of the ice, but until measurements are taken on the Moon this is the best assumption that can be made. Above the sublimation temperature of each component, the weighting is renormalized for the reduced mass. The heat capacity for the composite ice and for pure water are compared in Fig. 16. The integrated heat capacity of pure water is found to be always within 14% of the integrated heat capacity for composite ice. In the present accuracy of approximation, pure water's heat capacity can be used as adequate representation of the composite ice. The combined specific heat of regolith with 8.9%wt ice (now approximating it is all water) is shown in Fig. 17. The specific heat of water ice is roughly 3 times higher than the specific heat of dry lunar soil, but since it constitutes only 8.9%wt of the regolith it raises total heat capacity of the mixture by only about 29% at 40 K and about 16% at 200 K.

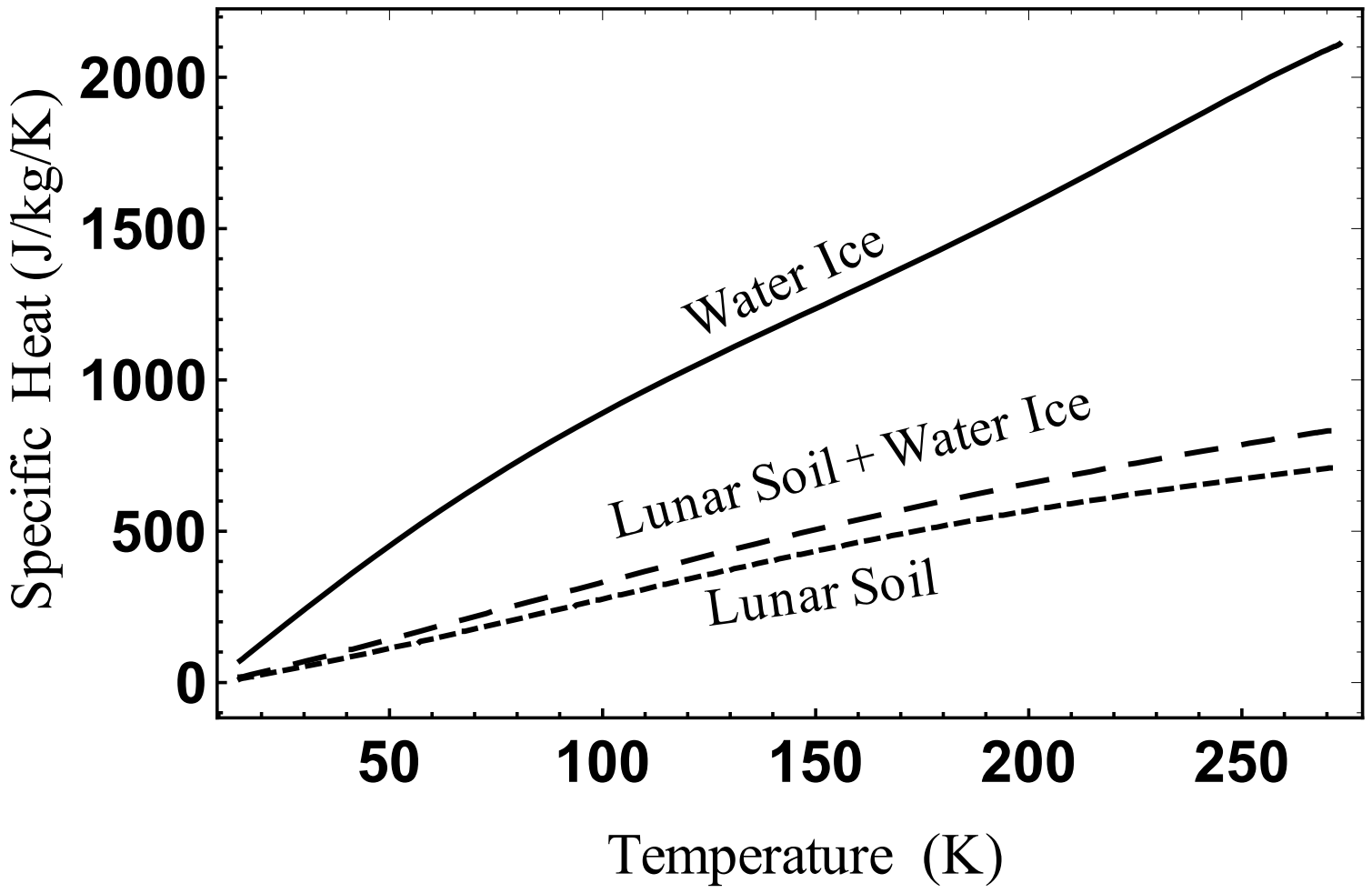

**Figure 17.** Thermal conductivity of water ice, lunar soil, and 8.9 %wt water ice in lunar soil.

## Phase Change of Ice

The sublimation of water ice on the Moon at temperatures below the triple point is treated by Andreas (2007) by relating the saturation vapor pressure of water ice, $e_{\mathrm{sat,i}}(T)$, which is a function of temperature *T*, to the sublimation rate $S_0$,

$$S_0 = e_{sat,i}(T)\left(\frac{M_W}{2\pi RT}\right)^{\frac{1}{2}} \tag{40}$$

in kg/m$^2$/s, where $M_W$ is the molecular weight of water and *R* is the universal gas constant. Kossiacki and Jacek (2014) modified this by subtracting the partial pressure of the vapor $P$ from the saturation pressure,

$$S_0 = \left[e_{sat,i}(T) - P\right]\left(\frac{M_W}{2\pi RT}\right)^{\frac{1}{2}} \tag{41}$$

When partial pressure reaches saturation pressure, then sublimation should cease. Here, the model will use the vapor pressure relationship provided by Murphy and Koop (2005),

$$e_{sat,i}(T) = 14050.7\, T^{3.53068}\, exp\left(-\frac{5723.265}{T} - 0.00728332\, T\right) \tag{42}$$

in Pa.

The free surface area of the ice where sublimation takes place depends on the physical state of the ice, whether it exists as large cobbles of ice surrounded by regolith fines, or as fine particles of ice comparable to the size of regolith particles intermixed with the mineral grains, or as a rind of ice coating the mineral grains, or as amorphous material residing in the pore spaces between grains, or as another form. How it is modeled depends on which physical state is assumed. One simple way to model the exposed surface area of the ice $\Delta s$ (in a cell toroidal of radius $r$ about the model's axis because this is an axisymmetric model) is $\Delta s = 2\pi r \Delta z\ \sigma$ where $\sigma$ is a parameter based on expected physical state of the ice informed by the geological model. The net sublimed mass during the $n^{\text{th}}$ timestep in cell location $(i, j)$ is therefore

$$m_{S,ij}^{n} = 2\pi r_{ij} \Delta z\ \sigma_{ij} \left[e_{\text{sat,i}}\left(T_{ij}^{n}\right) - P_{ij}\right] \left(\frac{M_W}{2\pi R T_{ij}^{n}}\right)^{\frac{1}{2}} \Delta t \tag{43}$$

The mass of water in the regolith's pore spaces in the toroidal cell could be modeled as,

$$m_{w,ij}^{0} = \pi\left(r_{ij}^{2} - r_{i,j+1}^{2}\right) \Delta z\ \varphi_{ij}\ \nu_{ij}\ \rho_{I} \tag{44}$$

Where $\varphi_{ij}$ is the fraction of the pore space filled by ice and $\rho_I$ is the density of the ice. A relationship is needed between $\sigma_{ij}$ and $\varphi_{ij}$ to represent the physical state of the ice. For now, the model uses the simplification $\sigma_{ij} = \varphi_{ij}$. Heat capacity is a simple scaling between ice and regolith heat capacities by the amount of each mass within a cell. Temperature in a cell may continue to rise even as ice sublimes until it reaches the triple point, because sublimation is a slow process at these temperatures and the system remains in non-equilibrium. In practical cases that have been modeled, the temperature never rose as high as the triple point. As sublimation occurs, the heat of fusion is subtracted from the internal energy of the cell and the temperature is lowered accordingly before time-stepping the model to calculate conduction of heat again. The gas and the solid components in each cell are assumed to have the same temperature.

## Release of Volatiles from Asteroid Regolith

Hydrated minerals in asteroid regolith will release their volatiles as a function of temperature. For testing an asteroid mining prototype, it was economically beneficial to use lower temperature materials for early tests so a lower temperature simulant was developed using primarily epsomite, because it will release most of its water of hydration below 300 ºC. Curves were obtained through Thermo-Gravimetric Analysis (TGA) to determine mass of released volatiles at each temperature increment.

Examples of these curves are shown in Fig. 18 for asteroid simulant UCF-CI-1 measured by Metzger et al. (2019), the Orgueil meteorite by King et al. (2015), and epsomite by Ruiz-Agudo et al. (2006).

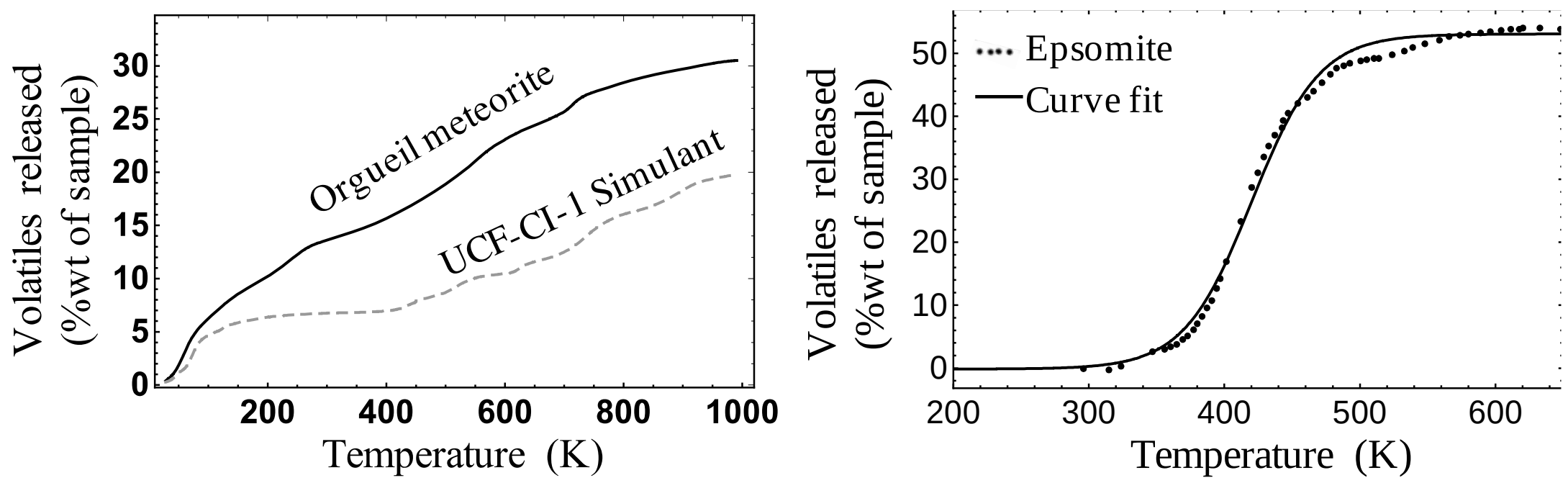

**Figure 18.** Thermo-Gravimetric Analysis (TGA): (Left) meteoritic and simulated asteroid materials; (Right) epsomite.

To model this, as a region in the model reaches a new high of temperature, the volatiles up to that temperature per the TGA curve are released as vapor, and the model remembers that no more volatiles will be released from that location until an even higher temperature is achieved. The appropriate amount of gas per the TGA curve is added to the gas already in the pore space at that location. Energy spent liberating the volatiles in each step is removed from the regolith appropriately in each time step. Thermal and gas diffusion are then iterated. The equation to fit epsomite (Fig. 18, Right) is,

$$w = \left[\tanh\left(\frac{T}{50} - 8.404\right) - \tanh\left(\frac{273.15}{50} - 8.404\right)\right] \times 26.61\%\text{wt} \qquad (45)$$

where $w$ is the weight percent (of a cell's material) that has sublimed, and $T$ is in kelvins.

The model does not modify the thermal conductivity of the solid fraction of the soil as mass is converted to vapor, although it should reduce that term because (especially with epsomite) a large fraction of the solid mass is lost, reducing the solid conduction contact network. That effect is offset by the increase in conductivity due to rising pore pressure as shown in Figure 13, but there are no empirical data at present to guide this improvement. Those experimental measurements and modeling are left to future work.

## Gas Diffusion

The model incorporates diffusion using the finite difference equations of Scott and Ko (1968). As vapor is evolved as described above the pressure differences drive it into neighboring cells. To couple the fast gas diffusion equations and the slow thermal diffusion equations while maintaining stability of the model, it was necessary to implement different time steps for each set of equations. Adaptive timesteps were implemented for the fast diffusion process, dividing each heat flow timestep into the minimum number of smaller timesteps necessary for stable solution of the gas diffusion equations. As pressure gradients increase, the gas diffusion timesteps become smaller. The resulting model is fast, allowing simulation of an hour-long physical test in just five or ten minutes on a standard laptop computer.

## 1D Thermal Model Validation

Only limited testing of the model has been performed. The following four cases demonstrate aspects of the thermal algorithms and the overall code structure with increasing complexity. The first case is one dimensional (1D) simulations of the lunar regolith heating and cooling in sunlight at various latitudes in comparison with Lunar Reconnaissance Orbiter (LRO) Diviner data per Fig. 9a of Vasavada, et al. (2012). The lunar surface albedo as a function of angle and other parameters choices by Vasavada et al. were intertwined with choices of thermal conductivity to make their model match lunar data sets. Vasavada et al. (2012) used $k = 0.6$ mW/m/K for the most porous soil at the lunar surface ($\nu_0 = 0.58$, bulk density $\rho_0 = 1300$ kg/m$^3$), asymptotically approaching $k = 7$ mW/m/K for the least porous soil at depth ($\nu_\infty = 0.42$, bulk density $\rho_\infty = 1800$ kg/m$^3$), with the porosity exponentially decaying as a function of depth, $z$,

$$\nu = \nu_0 - (\nu_0 - \nu_\infty)e^{-z/H} \tag{46}$$

where the "$H$ parameter" is the single remaining model parameter. Values of $H$ can be iterated until the model makes predictions that match observations of lunar surface temperature rising and falling as the Moon rotates in the sunlight. The value of $H$ is thus a proxy to characterize how rapidly the soil compactifies with depth at each location on the Moon in some averaged sense.

Following the choices of Vasavada et al., the thermal conductivity form in Eq. (15) becomes,

$$A = e^{-6.898+15.232(1-\nu)} \text{ (mW/m/K)}, \;\; \chi = e^{0.9933} \tag{47}$$

Ice content is set to zero. The specific heat is represented by Eq. (38). Simulations were performed for the Moon rotating in the sunlight over 37 lunations (months) to achieve steady state. The final lunation is shown in Fig. 19 for three cases: $H = 0.5$ cm (short dots, top curve, most compacted soil so highest thermal inertia), 3.5 cm (best fit, solid curve), and 30 cm (long dashes, bottom curve, loosest soil so least thermal inertia). The model was successful in predicting lunar temperatures indicating the model is structured correctly. Future work will use the improved parameterization of Eq. (25), which will make it necessary to determine how albedo and the other model parameters must be changed from the values of Vasavada et al. to match lunar measurements. This should produce improved characterization of $H$.

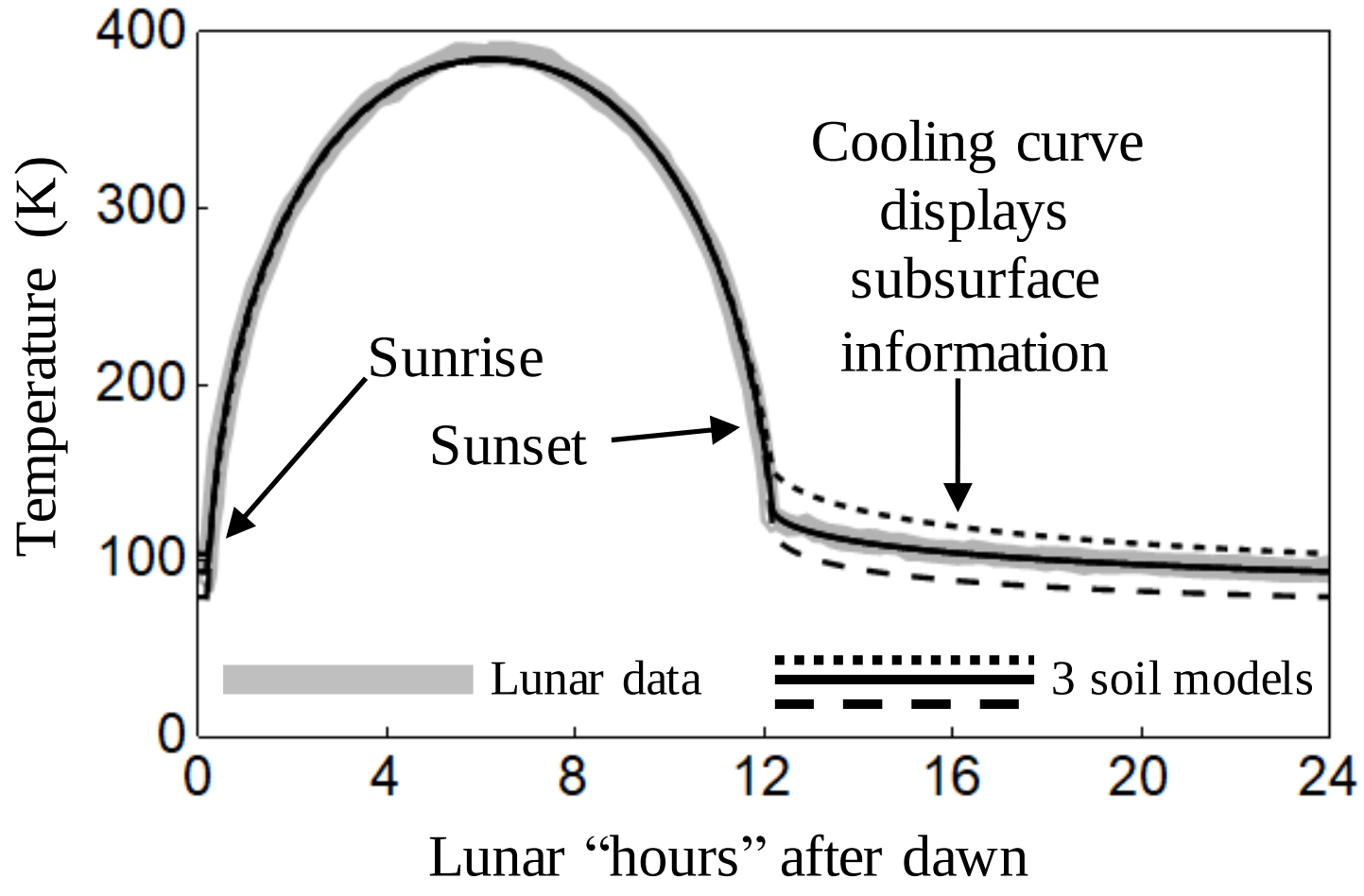

**Figure 19.** 1D modeling of the lunar case.

The second case is for equatorial conditions on asteroid 101955 rotating in sunlight. Fine tuning of the model has not been performed for the asteroid because adequate data sets from asteroids do not exist, but better data are expected soon from spacecraft missions currently in progress. Parameterization is therefore speculative. Many cases were modeled, and they produced similar results with differences that can be tested when the mission data become available. This particular case shown in Fig. 20 used a three-layer regolith model assuming the surface and deepest layers of the asteroid have identical properties while a layer with different properties exists between 0.5 and 6.0 cm depth. Theory says such an intermediate layer might form on asteroids by thermal cracking as the asteroid rotates in the sunlight while the uppermost layer loses the fines in the low gravity. The surface and deepest layers were assumed to be very porous with bulk density $\rho = 1300$ kg/m$^3$ while the intermediate layer was assumed to have $\rho = 2340$ kg/m$^3$. The specific heat was assumed the same as lunar soil in all layers per Eq. (38). The thermal conductivities of all three layers were assumed to follow Eq. (15). For the surface and deepest layers, parameter $A$ was estimated by taking the value of Vasavada et al. (2012) for the most porous lunar soil, $k$ =0.6 mW/m/K, then multiplying by the particle size factor suggested by Presley and Christensen (1997), $(D_{\text{asteroid}}/D_{\text{lunar}})^{0.5}$, where $D_{\text{asteroid}}$ represents average particle size of the asteroid regolith and $D_{\text{lunar}}$ represents average particle size of lunar soil. This could be interpreted as the expected larger contact patches because asteroid regolith is dominated by large gravel particles. $D_{\text{asteroid}}$ = 1.5 cm and $D_{\text{lunar}} = 60$ μm result in $A = 9.49$ mW/m/K. $\chi = 2.7$ is kept matching Vasavada, et al (2012). The thermal conductivity of the intermediate layer was assumed to have $A = 3.736$ mW/m/K, corresponding to an intermediate value between the most and least compacted lunar soil, and $\chi = 0.434$ for reduced radiative heat transfer due to reduced porosity. The simulation replicated the solar insolation conditions for Bennu and its approximately 4.3 hour rotation rate for 500 rotations to achieve steady state. The resulting range of temperatures shown in Fig. 20 correctly matches the range observed on Bennu as it rotates in the sun per Lauretta et al. (2015).

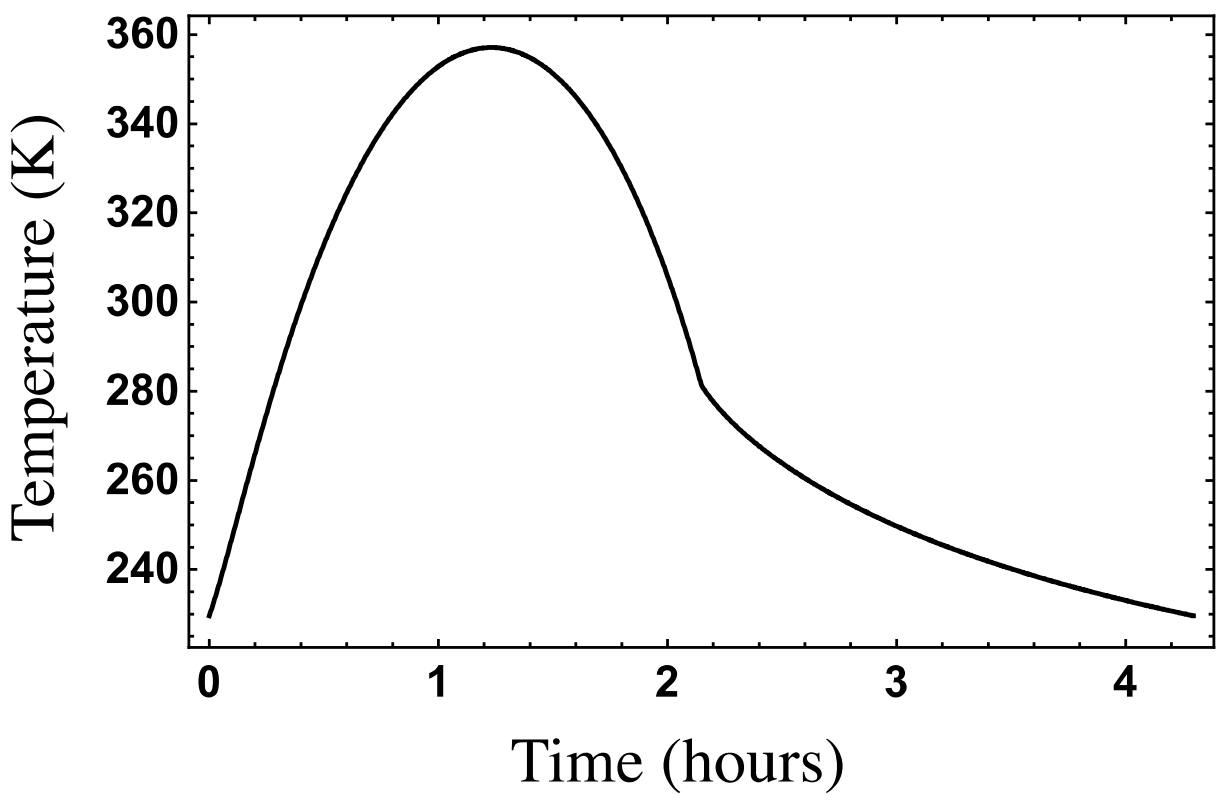

**Figure 20.** 1D modeling of asteroid 101955 Bennu

## 2D Axisymmetric Model Validation

The third case adds complexity by using the 2D axisymmetric version of the Crank Nicolson formulation while retaining the lunar soil property equations of the 1D model (following Vasavada et al.). It is a simulation for a drilling test in which a warm drill bit is embedded in soil that carries away its heat. The soil is in a tall, narrow, cylindrical container 14 cm in radius and 120 cm tall. The experiment is inside a warm vacuum chamber. This is the geometry of simulated tests done with a Honeybee Robotics drill at a NASA Glenn Research Center vacuum chamber where a liquid nitrogen bath kept the soil container at constant temperature (77 K) and removed heat from the soil conductively.

In these simulations, four different boundary temperatures (133 K, 153 K, 173 K, and 193 K) instead of the liquid nitrogen bath temperature were successively used to test how the Resource Prospector Mission drill bit could measure cooling rate while embedded in soil. The initial soil temperature before drilling was set to the boundary temperature. The soil model had no ice and was set to $\rho_0$ =1300 k/m$^3$ ($\nu_0 = 0.58$), $\rho_\infty$ =1950 k/m$^3$ ($\nu_\infty = 0.37$), $H$ = 5 cm, porosity following Eq. (46), heat capacity per Eq. (38), thermal conductivity per Eq. (15) parameterized by,

$$A = e^{-5.424+11.717(1-\nu)} \text{ (mW/m/K)}, \quad \chi = e^{0.9933} \tag{48}$$

Albedo and emission at the surface follow Vasavada et al. (2012). The vacuum chamber walls are set to 193 K for radiative heat transfer. The drill bit was initially held at temperature 213 K for 5,000 time steps (2 s each) while the soil came to equilibrium, shown in Fig. 21. Then it was allowed to cool to determine the cooling rate of the bit. The bit and drilling mechanism attached to its upper end were modeled for realistic thermal inertia. The model showed that the cooling process is so slow in lunar soil that it takes hours for soil around a warm drill bit to return to ambient temperature, matching the experimental observations. It is impractical for a lunar rover to pause its mission until the soil cools to take a subsurface temperature measurement. Fig. 22 shows that the cooling rate of the bit depends on the boundary temperature condition, which represents the original subsurface soil temperature before drilling. Thus, the lunar drill can quickly measure the cooling rate and continue its mission, relying on modeling to calculate the original subsurface temperature. The actual lunar case would be more complex than what was simulated here, because boundary temperatures (temperature asymptotically far from the drill) should vary with depth. Additional uncertainty will exist from compaction of the soil and ice content with depth. These additional unknowns will be informed in part by drill torque as it is inserted into the subsurface and by analysis of the cuttings for ice content as the drill brings the cuttings up to instruments at the surface. All these measurements would need to be analyzed to inform model parameterization. A future study could analyze accuracy of this overall method and its sensitivity on the several parameters.

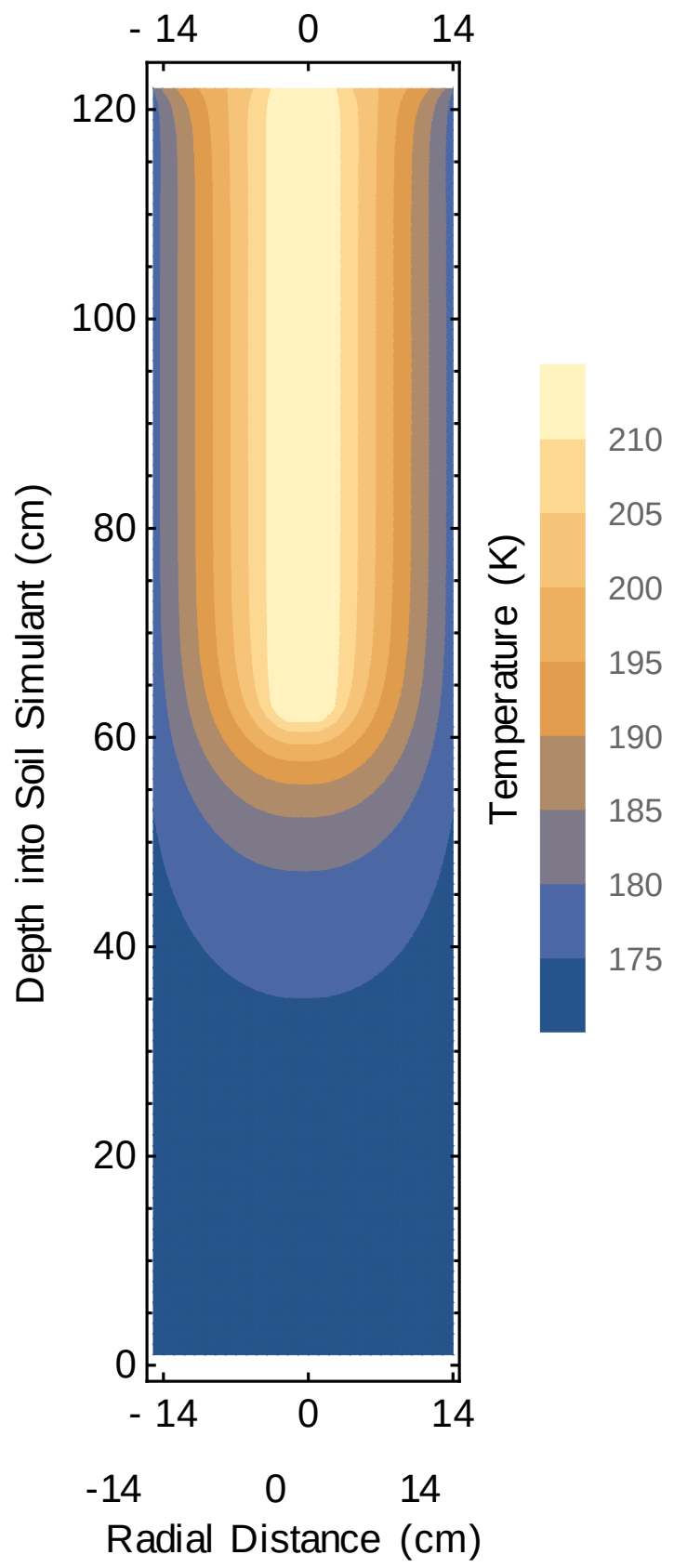

**Figure 21.** 2D axisymmetric simulation of warm drill bit in frozen lunar simulant. Lighter colors represent hotter soil.

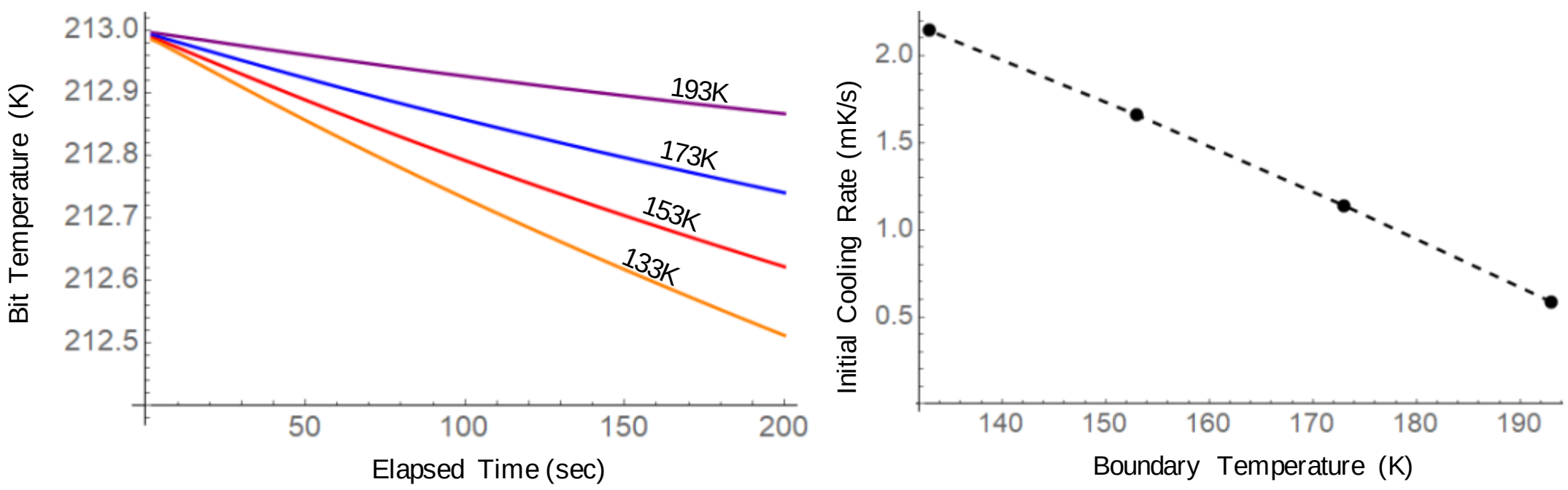

**Figure 22**: (Left) Bit temperature while cooling for four cases with different boundary temperatures 133K to 193K. (Right) Initial cooling rate of the bit versus boundary temperature.

## Thermal Extraction of Water from an Asteroid Simulant

The fourth case integrated the fully set of new constitutive equations given by Eqs. (36), (38), and (45) into the 2D axisymmetric Crank-Nicolson model. This was used to simulate the extraction of water from asteroid regolith by the WINE spacecraft coring device (Zacny et al., 2016; Metzger et al., 2016). The corer is a hollow tube with flutes on the exterior that enable it to drill into the subsurface, filling the hollow center of the tube with regolith. The corer walls are heated on the inside, while its layered insulating structure minimizes heat transfer to its exterior. Regolith increases in thermal conductivity as it warms, so the process becomes increasingly efficient. It releases volatiles according to the TGA curve, which further increases thermal conductivity.

The simulations were performed both for terrestrial test conditions with a 1 bar background pressure in the regolith and for space applications with the pores initially in vacuum. The soil begins at 272 K temperature and the soil container's boundaries are kept at 272 K throughout the simulation. In each timestep, 200 W thermal energy is delivered to the inside walls of the corer uniformly along its heated surface, then it diffuses from the tube through the soil.

Videos were created from the simulation data showing the resulting temperature and pressure fields in the regolith. Fig. 23 shows a series of snapshots of the temperature field in cross-section through the corer. The simulation demonstrated that corer design successfully keeps most of the thermal energy inside the interior although some energy leaks to the exterior. The videos show that the pressure builds up almost immediately then decays and the pressure field becomes more uniform. This decay is because the simulant's water of hydration becomes depleted. Fig. 24 shows a series of snapshots of the vapor pressure field. The semicircular pressure gradient at the top inside the corer is where the vapor diffuses to the collection tube located on the centerline. In the corresponding experiments, the tube leads to the cold trap where volatiles are frozen, keeping the tube at near vacuum conditions, but in the simulation the vapor that reaches the tube's entrance is simply accounted for then removed from the simulation to maintain the tube entrance at vacuum conditions. In the initial simulations a significant fraction of the vapor can be seen exiting the bottom of the corer rather than diffusing into the collection tube, reducing the system's mining efficiency. In this particular case the soil's initial temperature started near the triple point, so as it was warmed the vapor exiting the bottom of the corer did not freeze elsewhere in the soil but diffused to the soil's upper surface where it escaped into the surrounding vacuum.

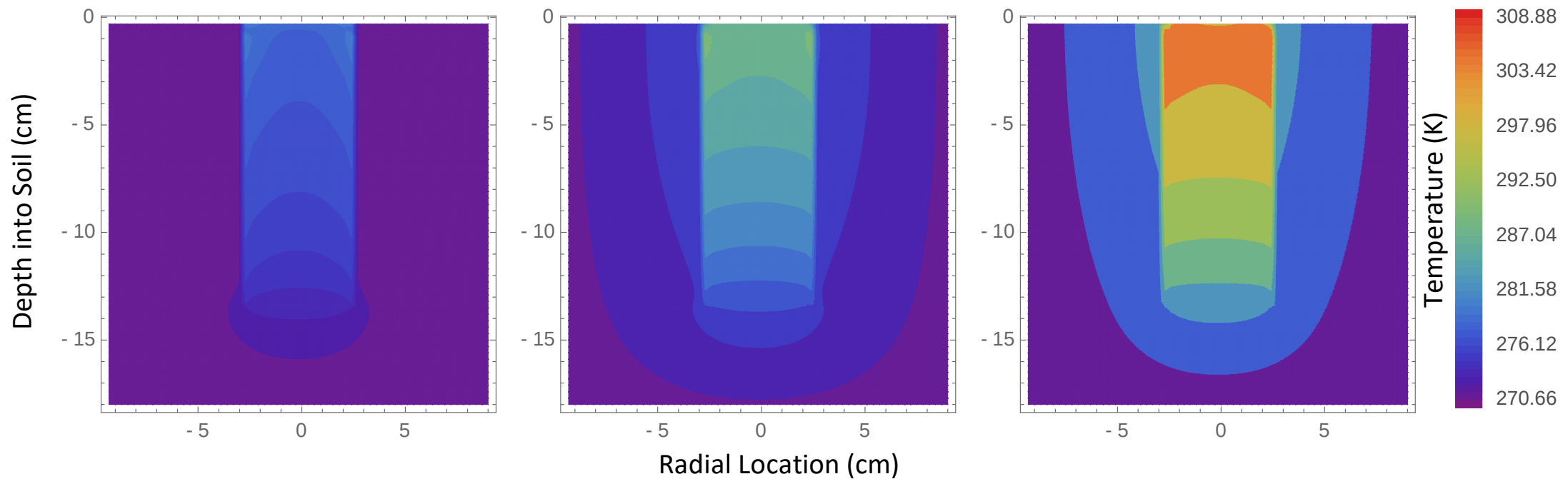

**Figure 23.** Temperature field in the Honeybee Corer at $t = 18$ s (left), $t = 90$ s (middle) and $t = 1080$ s (right).

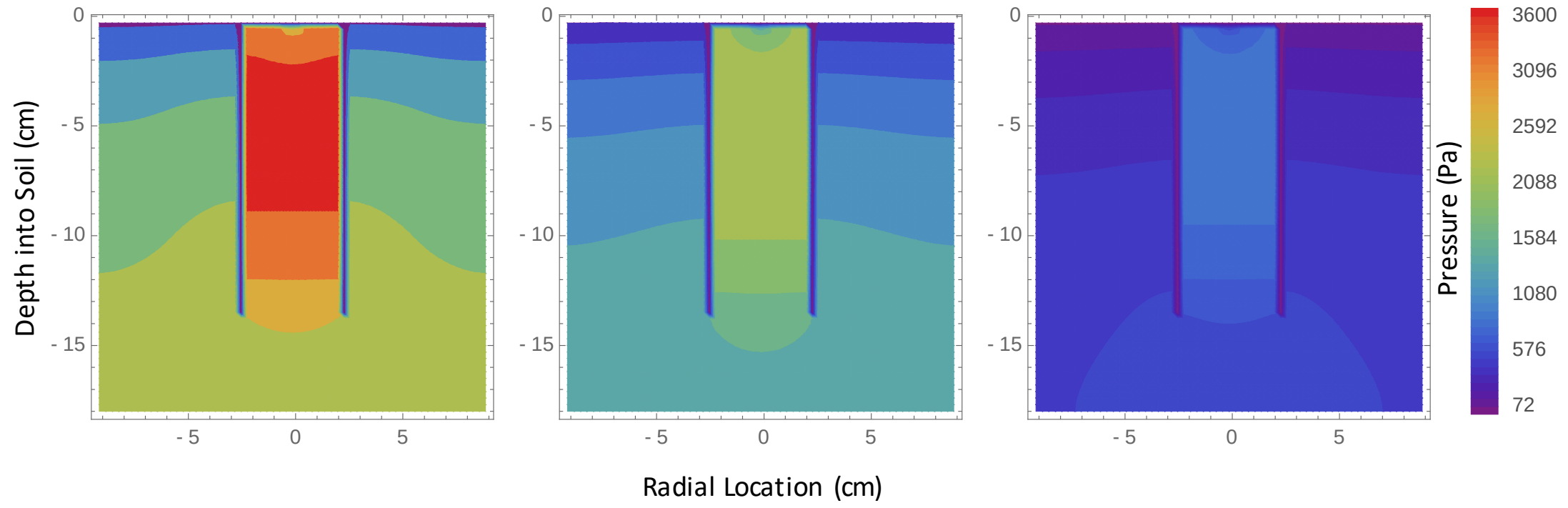

**Figure 24.** Pressure field in the Honeybee Corer driven fully into the soil at $t = 18$ s (left), $t = 90$ s (middle) and $t = 1080$ s (right).

To study how to capture a larger fraction of the vapor, additional simulations were performed where a gap was left between the top of the soil and the inside top of the coring tube. This can be achieved experimentally by not driving to corer all the way into the soil. This is the case shown in Fig. 25. The gap is so small it is not visible, but it is simulated by appropriate choice of model parameters so the gas can diffuse out from the soil into vacuum all along its top surface inside the corer. This produced a flat pressure gradient across the entire top of the soil inside the corer instead of the semicircular pressure gradient in Fig. 24. Comparing with Fig. 24, the vapor pressure outside the coring tube was reduced because vapor was transported upward through the corer more efficiently. This increased water capture by 520%. This illustrates how the modeling can be used to drive design of mining devices for improved performance in an extraterrestrial environment.

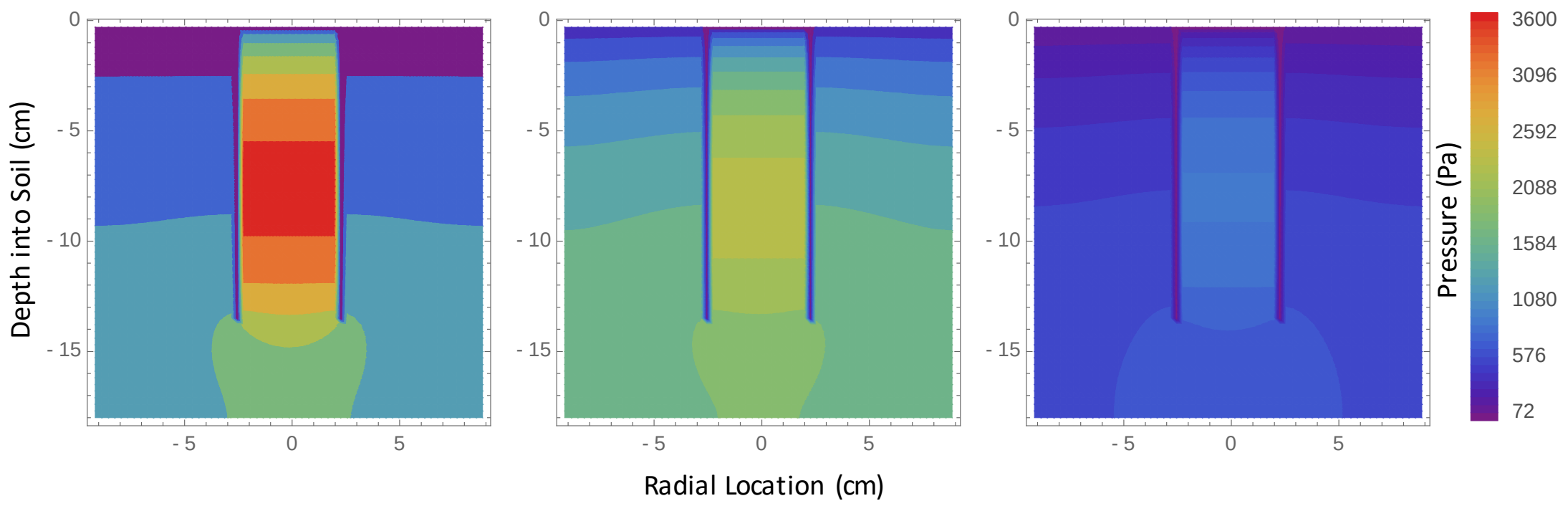

**Figure 25.** Pressure field in the Honeybee Corer leaving a gap at the top of the soil at $t = 18$ s (left), $t = 90$ s (middle) and $t = 1080$ s (right).

## CONCLUSIONS

Thermal volatile extraction modeling has been successfully developed for the 1D and axisymmetric 2D cases for asteroid and lunar regolith. The modeling includes parameterization for regolith thermal conductivity and heat capacity based on measurements of lunar soil samples, simulants, and terrestrial soil and ices. This is apparently the first time a soil constitutive model has successfully reconciled datasets for temperature, porosity, and gas pore pressure variables into a single equation. The model has been only partially tested. It produced excellent agreement with LRO Diviner data of the Moon and estimates of asteroid Bennu heating and cooling as they rotate in the sun. The 2D axisymmetric features have been demonstrated by simulating the Resource Prospector drill cooling after insertion into lunar soil. The model can also simulate the effects of ices upon the thermal conductivity and heat capacity of the lunar regolith. (For the asteroid case, ice is not expected as the volatiles are in the form of hydrated minerals.) The model is based on the assumption that lunar ice is crystalline rather than amorphous, which is supported by some data although the presence of amorphous phases cannot be ruled out. More work is needed to adapt the model to include the effects of amorphous ice. The model also successfully integrated equations for volatile release and gas diffusion along with the thermal diffusion equations, employing multiple, adaptive time steps to handle the different characteristic times of each part of the physics. The fully integrated model has been demonstrated for the case of a corer heating and extracting volatiles from asteroid regolith.

## DATA AVAILABILITY

Some or all data, models, or code generated or used during the study are available from the corresponding author by request: Mathematica notebook containing data from figures 1-8, 10-14, 16-20, and 23-25; Excel spreadsheet containing data from figures 9 and 15.

## ACKNOWLEDGEMENT

This work was directly supported by NASA SBIR contract no. NNX15CK13P, "The World is Not Enough (WINE): Harvesting Local Resources for Eternal Exploration of Space." This work was directly supported by NASA's Solar System Exploration Research Virtual Institute cooperative agreement award NNA14AB05A.

## NOTATION

The following symbols are used in this paper:

| | |
|---|---|
| $A, B$ | model fitting coefficients defined in the text; |
| $a, b$ | model fitting exponents defined in the text; |
| $\hat{a}, \hat{b}, \hat{c}$ | model fitting parameters defined in the text; |
| $C$ | soil heat capacity; |
| $c_1, \dots, c_7$, | model fitting exponents defined in the text; |
| $d$ | soil layer thickness; |
| $d_2, \dots, d_4$ | model fitting exponents defined in the text; |
| $e_{\mathrm{sat,i}}$ | saturation vapor pressure of water ice; |
| Exp | exponential function; |
| $i$ | index of model cell location in vertical direction; |
| $j$ | index of model cell location in radial direction; |
| $k$ | thermal diffusivity constant; |
| $k_1, \dots, k_9$ | model fitting exponents defined in the text; |
| $k_{\mathrm{Chen}}$ | thermal diffusivity values measured by Chen (2008) |
| $k_{\mathrm{PC}}$ | thermal diffusivity values measured by Presley and Christensen (1997) |
| Ln | natural logarithm function; |
| $m_{\mathrm{S}}$ | mass of ice sublimed; |
| $M_W$ | molecular weight; |
| $n$ | index of model timesteps; |
| $O(\,)$ | order of magnitude; |
| $P$ | pore gas pressure in the soil; |
| $r$ | location in radial direction; |
| $R$ | universal gas constant; |
| $S_0$ | sublimation rate of ice; |
| $S_{\mathrm{r}}$ | moisture saturation of soil; |
| $t$ | time; |
| $T$ | temperature; |
| $V$ | volume; |
| $w$ | weight percent sublimed; |
| $z$ | depth into the soil column; |
| $\alpha$ | model thermal parameter defined in the text; |
| $\Delta$ | model step difference in variable $r, z,$ or $t$; |
| $\delta_z^2, \delta_r^2$ | calculus operators defined in the text; |
| $\varepsilon$ | model fitting exponent defined in the text; |
| $\nu$ | soil porosity; |
| $\pi$ | pi (the number); |
| $\rho$ | bulk density of soil; |
| $\rho_I$ | density of water ice; |
| $\sigma$ | parameter to define surface area of ice in a model cell |
| $\varphi$ | fraction of soil's mass that is ice |
| $\chi$ | radiative transfer coefficient; |

**Appendix A.** Particle Samples Analyzed

| Source | Abbrev. | Mean Size ($\mu$m) | Range ($\mu$m) | Porosity or Grading | Bulk Density (kg/$m^3$) | Material |
|---|---|---|---|---|---|---|
| Chen (2008) | Chen | 320 [1] | Uniform | 0.396 | 1600 [2] | Quartz Sand |
| Chen (2008) | Chen | 320 [1] | Uniform | 0.423 | 1530 [2] | Quartz Sand |
| Chen (2008) | Chen | 320 [1] | Uniform | 0.457 | 1440 [2] | Quartz Sand |
| Chen (2008) | Chen | 320 [1] | Uniform | 0.490 | 1350 [2] | Quartz Sand |
| Chen (2008) | Chen | 570 [1] | Uniform | 0.434 | 1500 [2] | Quartz Sand |
| Chen (2008) | Chen | 570 [1] | Uniform | 0.472 | 1400 [2] | Quartz Sand |
| Chen (2008) | Chen | 570 [1] | Uniform | 0.509 | 1300 [2] | Quartz Sand |
| Chen (2008) | Chen | 570 [1] | Uniform | 0.547 | 1200 [2] | Quartz Sand |
| Chen (2008) | Chen | 120 [1] | Uniform | 0.434 | 1500 [2] | Quartz Sand |
| Chen (2008) | Chen | 120 [1] | Uniform | 0.472 | 1400 [2] | Quartz Sand |
| Chen (2008) | Chen | 120 [1] | Uniform | 0.509 | 1300 [2] | Quartz Sand |
| Chen (2008) | Chen | 120 [1] | Uniform | 0.547 | 1200 [2] | Quartz Sand |
| Chen (2008) | Chen | 310 [1] | Well Graded | 0.354 | 1710 [2] | Quartz Sand |
| Chen (2008) | Chen | 310 [1] | Well Graded | 0.396 | 1600 [2] | Quartz Sand |
| Chen (2008) | Chen | 310 [1] | Well Graded | 0.434 | 1500 [2] | Quartz Sand |
| Chen (2008) | Chen | 310 [1] | Well Graded | 0.472 | 1400 [2] | Quartz Sand |
| Presley and Christensen (1997) | PC | 805 [3] | 710 – 900 | 0.231 [4] | 2000 | Glass Spheres |
| Presley and Christensen (1997) | PC | 510 [3] | 500 – 520 | 0.308 [4] | 1800 | Glass Spheres |
| Presley and Christensen (1997) | PC | 262.5 [3] | 250 – 275 | 0.231 [4] | 2000 | Glass Spheres |
| Presley and Christensen (1997) | PC | 170 [3] | 160 – 180 | 0.346 [4] | 1700 | Glass Spheres |
| Presley and Christensen (1997) | PC | 154.5 [3] | 149 – 160 | 0.346 [4] | 1700 | Glass Spheres |
| Presley and Christensen (1997) | PC | 127.5 [3] | 125 – 130 | 0.423 [4] | 1500 | Glass Spheres |
| Presley and Christensen (1997) | PC | 95 [3] | 90 – 100 | 0.346 [4] | 1700 | Glass Spheres |
| Presley and Christensen (1997) | PC | 72.5 [3] | 70 – 75 | 0.423 [4] | 1500 | Glass Spheres |
| Presley and Christensen (1997) | PC | 27.5 [3] | 25 – 30 | 0.462 [4] | 1400 | Glass Spheres |
| Presley and Christensen (1997) | PC | 17.8 [3] | 15.6 – 20 | 0.538 [4] | 1200 | Glass Spheres |
| Presley and Christensen (1997) | PC | 13.3 [3] | 11 – 15.6 | 0.654 [4] | 900 | Glass Spheres |
| Fountain and West (1970) | FW | 49.5 [3] | 37 – 62 | 0.737 [5] | 790 | Crushed Basalt |
| Fountain and West (1970) | FW | 49.5 [3] | 37 – 62 | 0.707 [5] | 880 | Crushed Basalt |
| Fountain and West (1970) | FW | 49.5 [3] | 37 – 62 | 0.673 [5] | 980 | Crushed Basalt |
| Fountain and West (1970) | FW | 49.5 [3] | 37 – 62 | 0.623 [5] | 1130 | Crushed Basalt |
| Fountain and West (1970) | FW | 49.5 [3] | 37 – 62 | 0.567 [5] | 1300 | Crushed Basalt |
| Fountain and West (1970) | FW | 49.5 [3] | 37 – 62 | 0.500 [5] | 1500 | Crushed Basalt |
| Cremers and Birkebak (1971) | 12001,19 | 66 [6] | < 1000 [7] | 0.580 [8] | 1300 | Apollo 12 Lunar Soil |
| Cremers (1972) | 14163,133 | 68 [9] | < 1000 [10] | 0.645 [8] | 1100 | Apollo 14 Lunar Soil |
| Cremers (1972) | 14163,133 | 68 [9] | < 1000 [10] | 0.580 [8] | 1300 | Apollo 14 Lunar Soil |

Notes: [1] These mean sizes are D50 values scaled from Figure 1 in Chen (2008). [2] Calculated using the reported porosity and 2650 kg/m$^3$ for the mineral density of quartz. [3] Calculated as the mean of the end points of the reported range. [4] Calculated using the reported bulk density and 2.60 as the approximate specific gravity of the glass. [5] Calculated using the reported bulk density and 3.00 as the specific gravity of basalt. [6] Average of measurements reported in Meyer (2011a). [7] From Meyer (2011a). [8] Calculated using the reported bulk density and 3.10 as the mean specific gravity of lunar soil. [9] Average of measurements reported in Meyer (2011b). [10] From Meyer (2011b)